\documentclass[twocolumn, trackchanges]{aastex631}
\usepackage{amsmath}
\usepackage {graphicx}
\usepackage{ulem}
\usepackage{savesym}
\savesymbol{tablenum}
\usepackage{siunitx}
\restoresymbol{SIX}{tablenum}
\usepackage{soul}

\usepackage{breqn}
\usepackage[caption=false]{subfig}
\usepackage{chngcntr}

\definecolor{my_color}{HTML}{3a18b1}

\definecolor{new_color}{HTML}{000000}
\definecolor{new_black}{HTML}{000000}

\newlength{\wdth}

\usepackage{xfrac}
\usepackage{xcolor}
\usepackage{subfig}
\usepackage{mdframed}

\newcommand\bedit[1]{\textcolor{new_black}{#1}}
\shorttitle{Classifying X-Ray Binaries Using ML }
\shortauthors{de Beurs et al.}

\begin{document}

%

\title{A Comparative Study of Machine Learning Methods for X-ray Binary Classification \\
}

\correspondingauthor{Z L.\ de Beurs}
\email{zdebeurs@mit.edu}


\author[0000-0002-7564-6047]{Zoe L. de Beurs}
\affiliation{Department of Earth, Atmospheric, and Planetary Sciences \\
Massachusetts Institute of Technology \\
77 Massachusetts Ave, Building 54 \\
Cambridge, MA 02139, USA}
\affiliation{NSF Graduate Research Fellow}

\author[0000-0002-2413-9301]{N.\ Islam}
\affiliation{Center for Space Science and Technology, University of Maryland  \\
Baltimore County, 1000 Hilltop Circle \\
Baltimore, MD 21250, USA}
\affiliation{X-ray Astrophysics Laboratory, NASA Goddard Space Flight Center  \\
Greenbelt, MD 20771, USA}

\author[0000-0002-9630-7917]{G.\ Gopalan}
\affiliation{Statistics Department, California Polytechnic State University  \\
San Luis Obispo, CA 93407, USA}

\author[0000-0002-7521-9897]{S. D.\ Vrtilek}
\affiliation{Center for Astrophysics \textbar\ Harvard \& Smithsonian  \\
60 Garden St. \\
Cambridge, MA 02138, USA}

\begin{abstract}
X-ray Binaries (XRBs) consist of a compact object that accretes material from an orbiting secondary star. The most secure method we have for determining if the compact object is a black hole is to determine its mass:  this is limited to bright objects, and requires substantial time-intensive spectroscopic monitoring. With new X-ray sources being discovered with different X-ray observatories, developing efficient, robust means to classify compact objects becomes increasingly important. We compare three machine learning classification methods (Bayesian Gaussian Processes (BGP), K-Nearest Neighbors (KNN), Support Vector Machines (SVM)) for determining the compact objects as neutron stars or black holes (BHs) in XRB systems. Each machine learning method uses spatial patterns which exist between systems of the same type in 3D Color-Color-Intensity diagrams. We used lightcurves extracted using six years of data with MAXI/GSC for 44 representative sources. We find that all three methods are highly accurate in distinguishing pulsing from non-pulsing neutron stars (NPNS) with 95\% of NPNS and 100\% of pulsars accurately predicted.  All three methods have high accuracy distinguishing BHs from pulsars (92\%) but continue to confuse BHs with a subclass of NPNS, called the Bursters, with KNN doing the best at only 50\% accuracy for predicting BHs.  The precision of all three methods is high, providing equivalent results over 5-10 independent runs. In a future work, we suggest a fourth dimension be incorporated to mitigate the confusion of BHs with Bursters.  This work paves the way towards more robust methods to efficiently distinguish BHs, NPNS, and pulsars.
\end{abstract}





\section{Introduction} \label{sec:intro}
X-ray binaries (XRBs) consist of a compact object accreting matter from a main sequence or supergiant companion star, orbiting the common center of mass.  They are often identified as Low-Mass X-Ray Binaries (LMXBs, companion star mass $\leq$ 1 \(M_\odot\)); and High-Mass X-Ray Binaries (HMXBs, companion star mass $\geq$ 10 \(M_\odot\)).  In general the compact object accretes matter through stellar wind capture in HMXBs and through Roche-lobe overflow in LMXBs.  Both HMXBs and LMXBs can contain a black hole (BH), a non-pulsing neutron star (NPNS), or a pulsar as the compact object. They are further distinguished by the presence or absence of pulsations, jets, bursts, spectral characteristics that vary over time, and variations in luminosity \citep{Paradijs}.

The most reliable means of determining the nature of the compact object is through searching for the presence of a surface (pulsations, cyclotron resonance scattering features, thermonuclear bursts, etc.), or by determining the mass of the compact object using radial velocity measurements. Overall, very few reliable methods exist to determine the nature of the compact object in an \bedit{XRB}. In addition, these methods are limited to bright objects, requiring extensive monitoring by various ground based and space based telescopes.  Although \bedit{hundreds of XRB candidates have been identified in our Galaxy \citep{2006Liu, 2007Liu}} since their first discovery over 60 years ago, we still lack a straightforward method to determine the nature of the compact objects.  


\begin{figure*}[h!tp]
\gridline{\fig{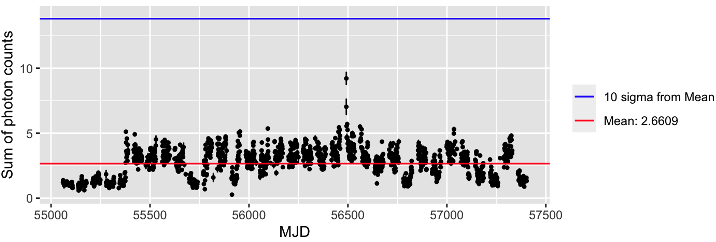}{0.53\textwidth}{(a) BH Cyg X-1}
          \fig{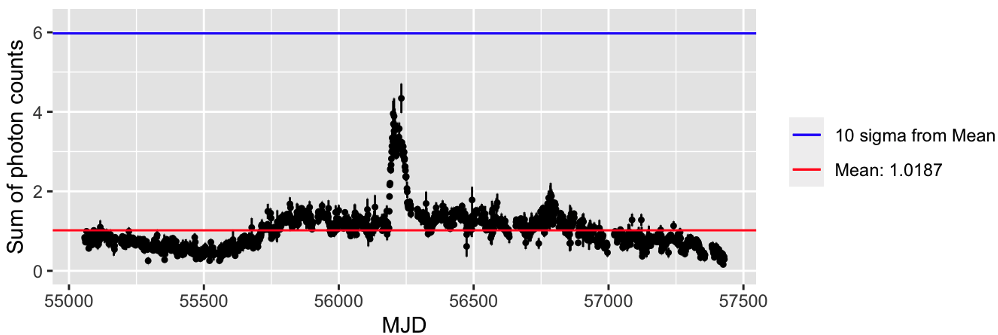}{0.53\textwidth}{(b) Burster GX3+1}
          }
\gridline{\fig{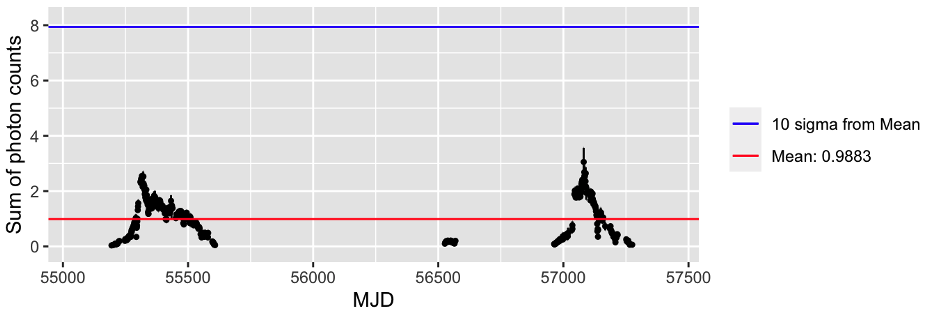}{0.53\textwidth}{(c) BH GX339-4}
          \fig{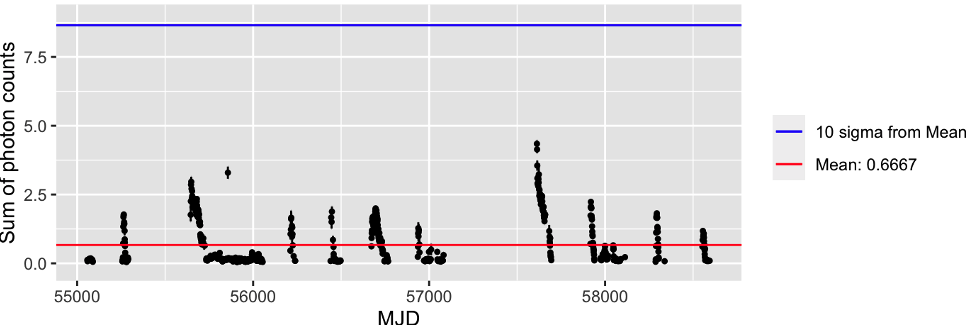}{0.53\textwidth}{(d) Burster 4U1608-52}
          }
\gridline{\fig{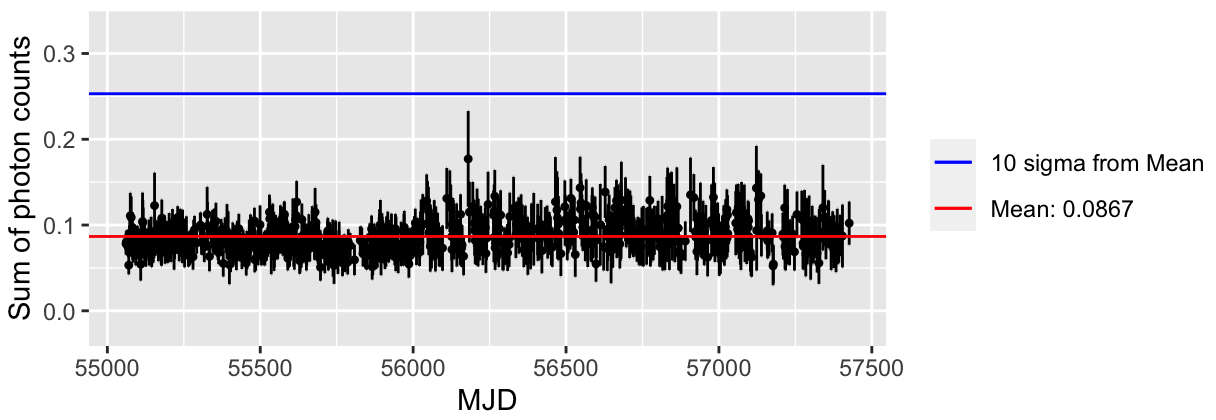}{0.54\textwidth}{(e) BH LMC X-1}
          \fig{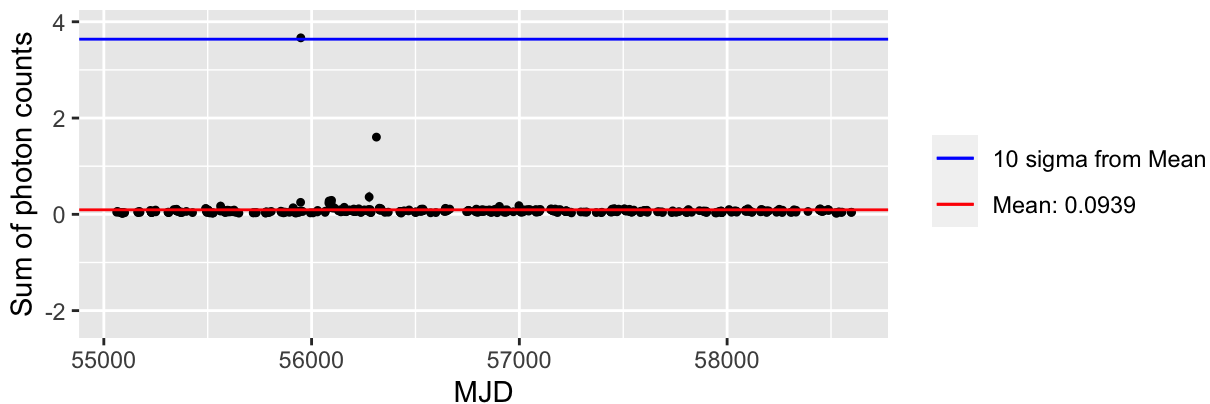}{0.53\textwidth}{(f) Burster 4U1916-053}
          }    
\caption{\footnotesize{Example lightcurves of BHs (a, c, e) and Bursters (b, d, f) showing the effect of the 10 sigma cutoff. The lightcurves are binned by 1 day.}
\label{fig:10sigma}}
\end{figure*}

\begin{deluxetable*}{lccc}[t!]
\LARGE
\tabletypesize{\footnotesize}
\tablenum{1}
\tablecaption{X-ray binary sources from MAXI data\label{table:1}}
\tablehead{
Source & Compact &
Source & Number of    \\
Name & Object Type &
class & Significant points  \\
}
\startdata
LMCX-3           & BH      & HMBH, persistent  & 765           \\
LMCX-1           & BH      & HMBH, persistent  & 821           \\
MAXIJ1535-571    & BH Candidate  & LMXB, transient   & 306           \\
4U1630-47           & BH Candidate  & LMXB, transient   & 679           \\
GX339-4          & BH      & LMBH, transient   & 502           \\
GRS1739-278        & BH Candidate  & LMXB, transient   & 1327          \\
H1743-322           & BH Candidate  & LMXB, transient   & 224           \\
MAXIJ1820+070    & BH Candidate  & LMXB, transient   & 158           \\
GRS1915+105          & BH      & LMBH, transient   & 1782          \\
CygX-1           & BH      & HMBH, persistent  & 1456          \\
4U1957+115       & BH Candidate  & LMXB, transient   & 1298          \\
CygX-3           & BH Candidate  & HMXB,  persistent & 1284          \\
\hline
H0614+091        & Non-Pulsing Neutron Star & LMNS, Burster     & 2176          \\
4U1254-690       & Non-Pulsing Neutron Star & LMNS, Burster     & 2915          \\
CirX-1            &  Non-Pulsing Neutron Star  &  LMNS, Burster      &  326            \\
4U1608-52        & Non-Pulsing Neutron Star & LMNS, Burster     & 709           \\
ScoX-1           & Non-Pulsing Neutron Star & LMNS   & 2059          \\
H1636-536         &  Non-Pulsing Neutron Star  &  LMNS, Burster      &  2228            \\
4U1700-37         &  Non-Pulsing Neutron Star  &  HMNS    &  1152            \\
GX349+2           &  Non-Pulsing Neutron Star  &  LMNS    &  2375            \\
4U1705-44         &  Non-Pulsing Neutron Star  &  LMNS    &  2222            \\
GX9+9 & Non-Pulsing Neutron Star & LMNS   & 2066          \\
GX3+1  &  Non-Pulsing Neutron Star  &  LMNS, Burster      &  1862            \\
GX5-1 & Non-Pulsing Neutron Star & LMNS   & 2659          \\
GX9+1  &  Non-Pulsing Neutron Star  &  LMNS    & 2618            \\
GX13+1           & Non-Pulsing Neutron Star & LMNS   & 2082          \\
GX17+2            &  Non-Pulsing Neutron Star  &  LMNS, Burster      &  2134            \\
SerX-1           & Non-Pulsing Neutron Star & LMNS, Burster     & 1863          \\
HETEJ1900.1-2455  &  Non-Pulsing Neutron Star  &  LMNS, Burster      &  1031            \\
AqlX-1           & Non-Pulsing Neutron Star & LMNS, Burster     & 274           \\
4U1916-053       & Non-Pulsing Neutron Star & LMNS, Burster     & 565           \\
CygX-2           & Non-Pulsing Neutron Star & LMNS   & 1446          \\
\hline
SMCX-1           & Pulsar  & HMNS   & 978           \\
LMCX-4           & Pulsar  & HMNS   & 192           \\
1A0535+262       & Pulsar  & HMNS   & 124           \\
VelaX-1          & Pulsar  & HMNS   & 686           \\
GROJ1008-57      & Pulsar  & HMNS   & 233           \\
CenX-3           & Pulsar  & HMNS   & 1187          \\
GX301-2          & Pulsar  & HMNS   & 192           \\
4U1538-52        & Pulsar  & HMNS   & 193           \\
4U1626-67        & Pulsar  & LMNS   & 1202          \\
HerX-1            &  Pulsar   &  IMNS    &  366             \\
OAO1657          & Pulsar  & HMNS   & 157           \\
4U1822-37        & Pulsar  & LMNS   & 919          \\  
\enddata
\tablecomments{The classifications for each of these sources are from \cite{2001Liu, 2006Liu, 2007Liu, 2010ApJ...718..488S, 2019MNRAS.487..928S, 2020ApJ...893L..37T}.
\vspace{-0.1in}}
\end{deluxetable*}

An important means of studying their spectral characteristics is through the use of color-color (CC) and color-intensity (CI) diagrams.  X-ray colors are defined as a ratio of photon counts in two X-ray energy bands. The information in CC diagrams captures the spectral states of XRBs while CI diagrams depict the variations in intensity over time.  For example, NPNS systems are classifed as Z or Atoll sources based on the shape they trace out in CC plots  \citep{1989A&A...225...79H}. Z sources trace out a Z-shape in X-ray color-color diagrams while Atoll sources trace out banana-shaped or circular structures. However, it has been shown that the same source can exhibit both geometric patterns, depending on the mass accretion rate \citep{Homan_2010, 2015ApJ...809...52F}. Often, different classes of XRBs also occupy overlapping regions in these 2-dimensional representations (CC, CI).

In 2013, \citeauthor{2013MNRAS.428.3693V} (hereafter VB13) proposed a 3-dimensional representation (color-color-intensity; CCI) of XRBs which placed different classes into geometrically different regions, providing a model-independent means of separating the types.  As a step toward understanding the physical mechanisms behind the separation seen by VB13,  \citet[hereafter GP15]{2015ApJ...809...40G} developed a probabilistic (Bayesian) model which uses a supervised learning approach (unknown classifications are predicted using known classifications), to quantify the accuracy of predicting the type of an unknown X-ray binary using the VB13 representation.  

\citet{pattnaik2021machine} tested 6 machine learning methods (ML) for their ability to accurately classify the compact objects in LMXBs, with CC and CI diagrams using data from the Proportional Counter Array (PCA) on RXTE \citep{1994ITNS...41.1343G}. They found that the Random Forest (RF) and K-nearest Neighbors (KNN) methods gave the highest accuracies and specifically evaluated the performance of the RF.  They found 87$\%$ accuracy of predictions for observations with SNR between 100-1000; for lower SNR data they achieved 58$\%$ accuracy.

\begin{figure*}[h!tp]
\centering
\epsscale{1}
\plotone{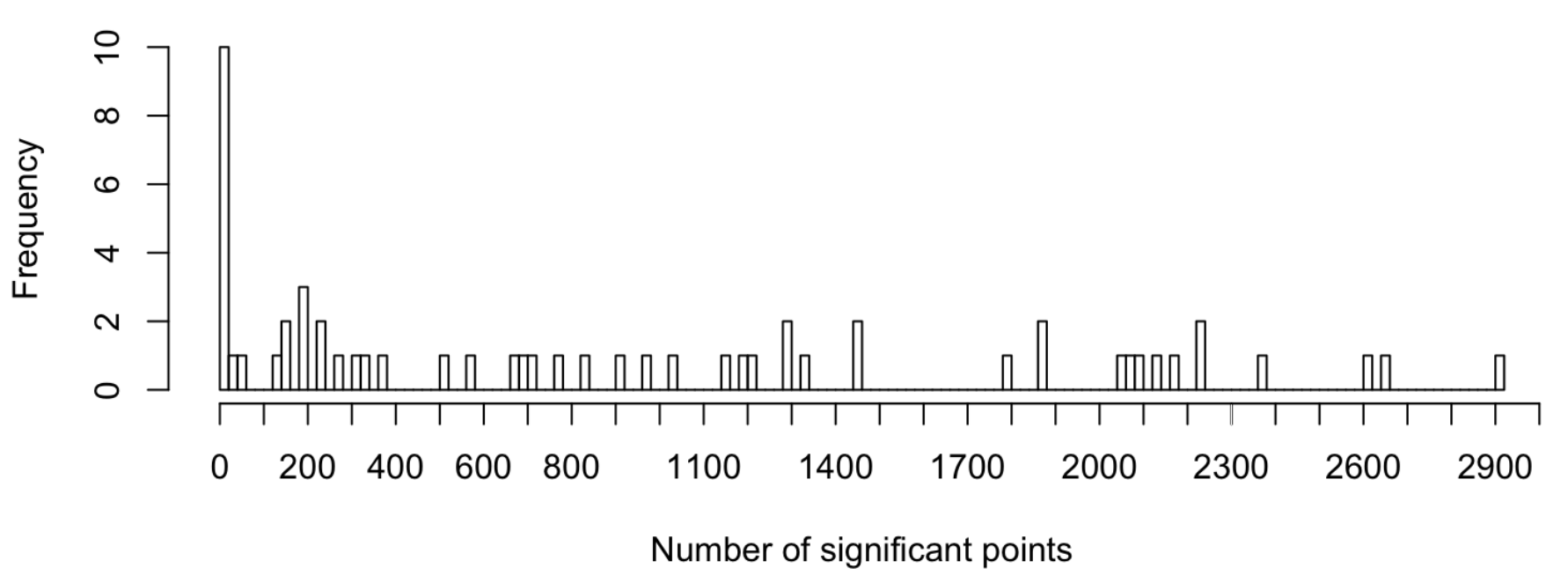}
\caption{\footnotesize{Histogram of the number of significant points for all sources. The frequency is the number of sources with this number of significant points.  There is a clear cut around 100 points: 10 sources have 0 cts, and 2 are below 50cts, whereas all the others have 120 cts or more.}
\label{fig:100pt}}
\end{figure*}

\begin{figure}[htp]
\centering
\epsscale{0.83}
\plotone{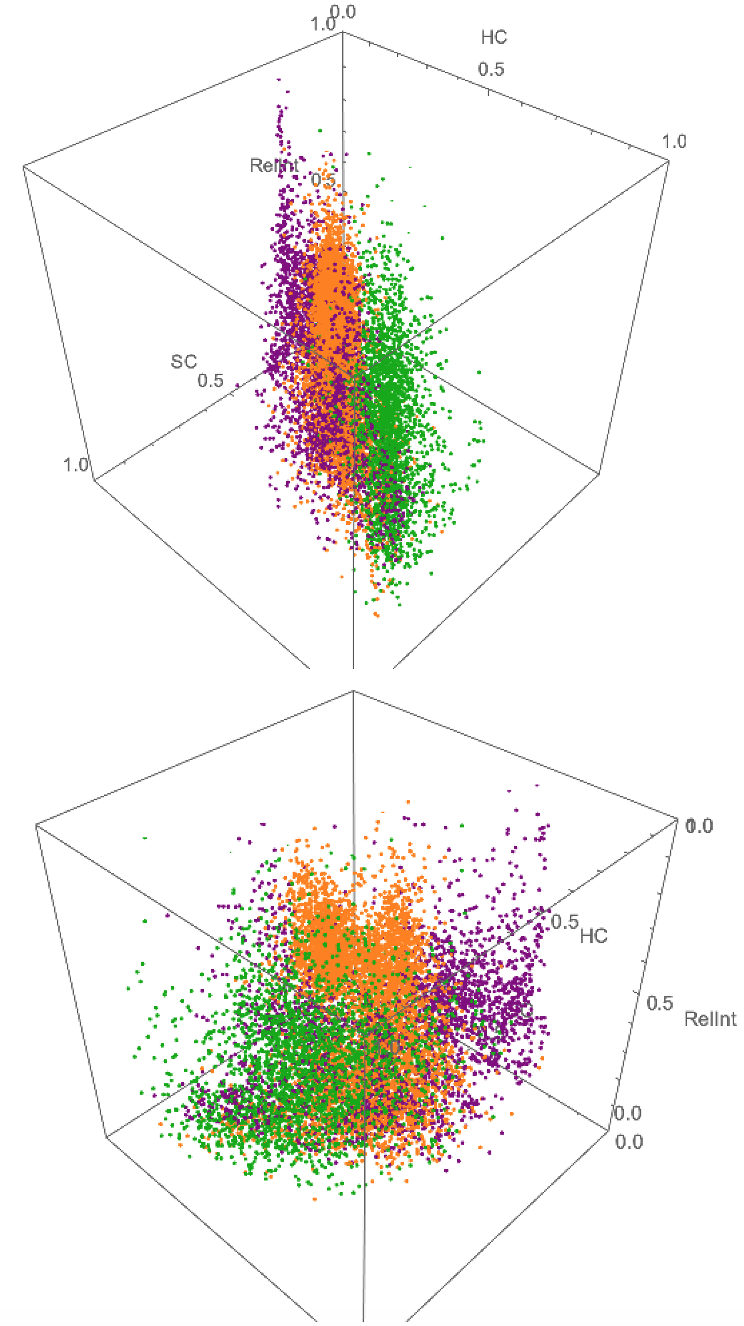}
\caption{\footnotesize{Color-Color-Intensity (CCI) diagrams of 12 BH (purple), 20 NPNS (orange), and 12 pulsar sources (green) from two different angles ($35^{\circ}$, $145^{\circ}$). Each point represents two X-ray colors and the corresponding intensity over one day for a given XRB.}
\label{fig:CCI}}
\end{figure}

\begin{figure}[htp]
\centering
\epsscale{0.79}
\plotone{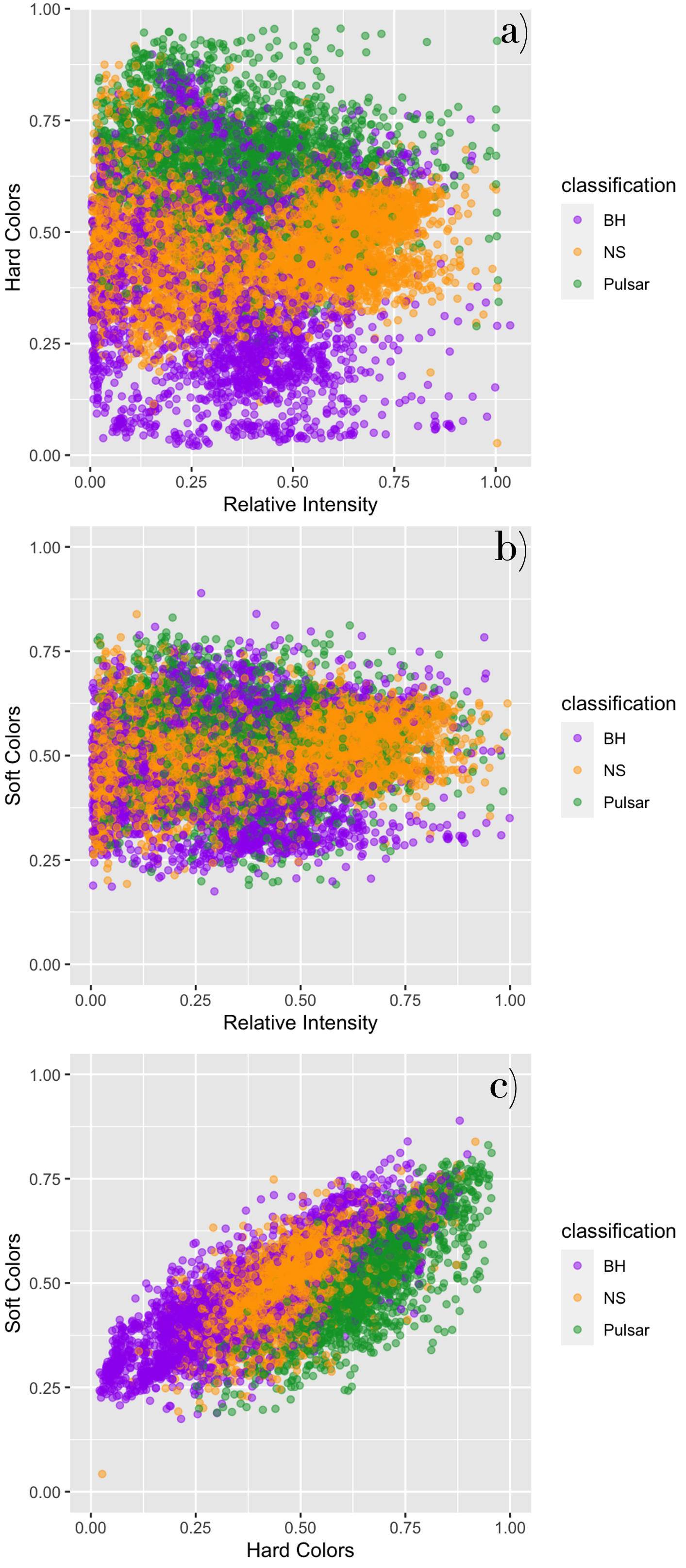}
\caption{\footnotesize{Two-dimensional projections of Color-Color-Intensity Diagrams of all 44 XRB sources. (a) Relative Intensity vs Hard Color for BHs (purple), NPNSs (orange), and pulsars (green). (b) Relative Intensity versus Soft Color for BHs, NPNSs, and pulsars.  (c) Hard Colors versus Soft Colors for BHs, NPNSs, and pulsars.}
\label{fig:CCI_2D}}
\end{figure}

\begin{figure*}[h!tp]
\centering
\epsscale{1.15}
\plotone{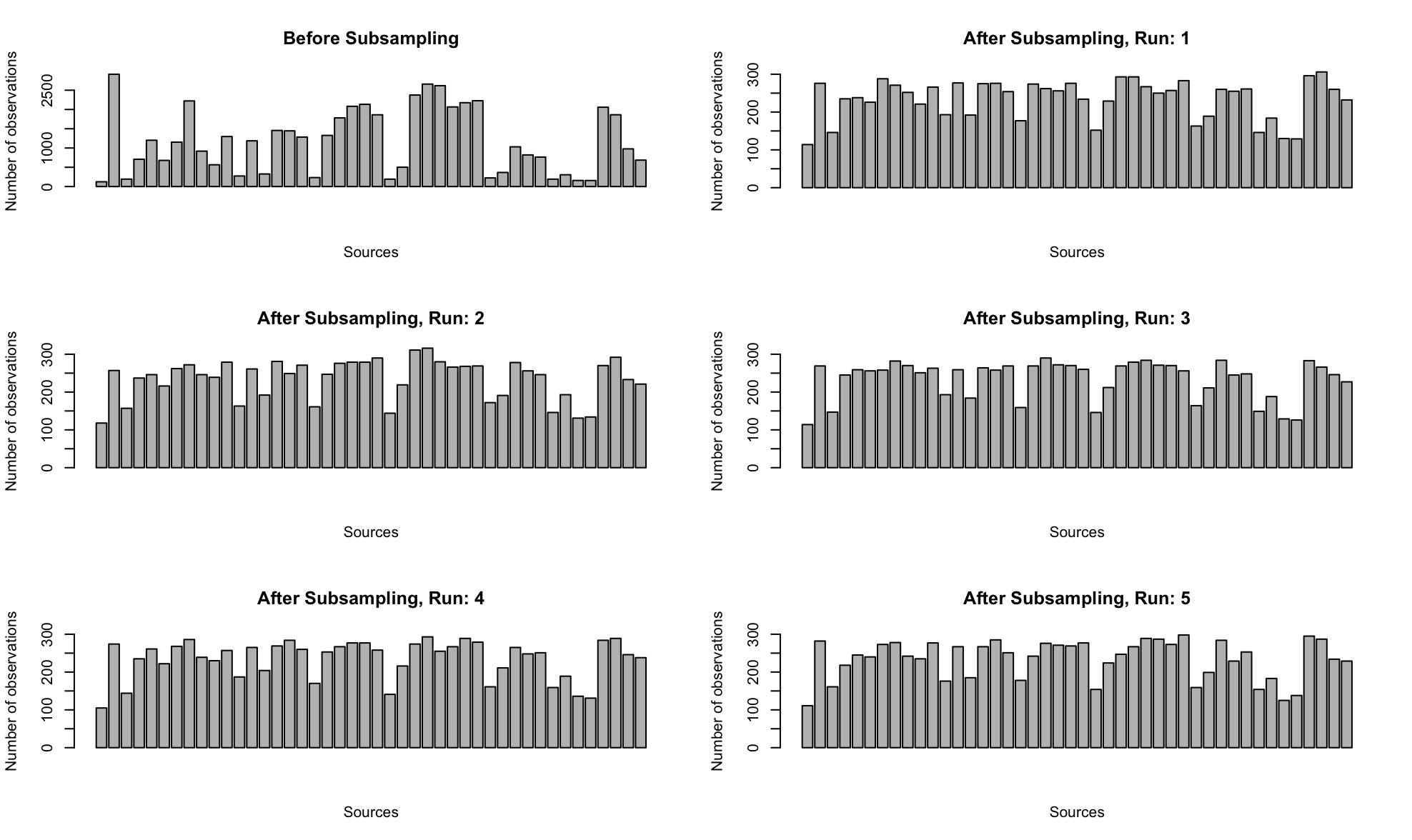}
\caption{\footnotesize{Distribution of the number of observations per source and five examples of  distributions after subsampling.  These were used to test BGP runs.  These and an additional five subsamples (Shown in \textit{Figure \ref{fig:subsamp2}}) were used to test KNN and SVM runs.}
\label{fig:subsamp}}
\end{figure*}
 
In this paper, we compare three ML techniques to determine which provides the most efficient and accurate means of identifying the nature of the compact object in both LMXBs and HMXBs.  We used the 3D representation of data introduced by VB13.  As demonstrated by  \citet{2021NewA...8501514I}, CCI has the advantage over CC and CI individually, in that the geometric patterns it produces translate consistently to data from different instruments.  The ML techniques we use, Bayesian Gaussian Process similar to GP15, KNN as used by \citet{pattnaik2021machine}, and Support Vector Machines (SVM), are widely used and particularly suitable for capturing spatial patterns in three dimensional data.  We use data from the Monitor of All Sky X-ray Image \citep[MAXI;][]{matsuoka} in the energy bands that \citet{2021NewA...8501514I} demonstrated most clearly show the separation of systems containing different types of compact object.  The advantage of all sky monitors such as RXTE/ASM or MAXI is that while RXTE/PCA had only pointed observations in time limited windows for specific behaviors of the system all sky monitors have long term monitoring that covers all stages of outbursts, states, and transitions.

This paper is organized as follows. In Section 2, we describe the observations from MAXI.  In Section 3, we describe the mathematical foundations of the ML algorithms implemented. In Section 4, we describe how these ML methods were applied to our observations. In Section 5, we present our results.  In Section 6, we compare the computational efficiency of the different methods.  In Section 7, we provide a summary and conclusions.

\section{Description of the data}
\subsection{Observations}
All the data used in this paper were obtained with the Gas Slit Camera \citep[GSC;][]{2002SPIE.4497..173M,2011PASJ...63..397T} on board the Monitor of All sky X-ray Image \citep[MAXI;][]{matsuoka}.  In operation since 2009, MAXI is the first astronomical mission to be operated on the International Space Station \citep[ISS;][]{matsuoka,10.1093/pasj/63.sp3.S635}. MAXI has higher sensitivity and higher energy resolution than any other all-sky-X-ray-monitors flown to date \citep{2002SPIE.4497..173M,2011PASJ...63..397T}. GSC has a 1 day sensitivity of 9 mCrab (3$\sigma$) compared to 15 mCrab with RXTE/ASM \citep{1996ApJ...469L..33L} and 16 mCrab with Swift/BAT  \citep{2013ApJS..209...14K}. The GSC on MAXI covers the energy band from 2 to 30 keV and contains six units of Xe-gas proportional counters which are assembled to cover wide fields of view of  1.$^\circ$5 x 160$^\circ$ \citep{10.1093/pasj/63.sp3.S635}. In addition, MAXI provides on-demand processing to extract lightcurves in user-specified energy bands\footnote{\url{http://maxi.riken.jp/mxondem}}. We specify the energy bands 2-3 (Low), 3-5 (Medium), and 5-12 keV (High) which are close to those used by VB13 and GP15 and demonstrated by \citet{2021NewA...8501514I} to be effective in separating classes of XRBs. 

In this paper, the soft colors (SC) are defined as 
\begin{equation}
    \text{SC = (Medium-Low)/(Medium+Low)}
\end{equation}
where Low and Medium refer to the counts in the 2-3 keV and 3-5 keV energy bands respectively. The hard colors (HC) are defined as 

\begin{equation}
    \text{HC = (High-Low)/(High+Low)}
\end{equation}

where High refers to the counts in the 5-12 keV energy band. \citet{park_2006} found that using this fractional difference ratio works well both in high-count and low-count regimes, whereas simple ratios or logarithms of ratios tend to fail in low-count regimes. \bedit{After we compute SC and HC, we rescale them to match the scale of the relative intensity, which ranges from 0 to 1. }

The relative intensity is computed by summing the counts in each of the energy bands (2-3 keV, 3-5 keV, 5-12 keV) and then normalizing each source by the average of the top 0.01\% of the counts to ensure the relative intensity is scaled from 0 to 1. The relative intensity (RelInt) can thus be written as
\begin{equation}
    \text{RelInt} = \frac{\text{Low + Medium + High}}
    {\text{99.99 \% percentile of the counts}}
\end{equation}

We only used detections with at least 3$\sigma$ significance for each XRB observed by MAXI. For a datapoint to be statistically valid, we require that it has at least 3$\sigma$ significance in the sum of the counts in the three energy bands. We achieve this by requiring that the counts in the individual energy bands are detected at the accuracy of $\sqrt{3\sigma}$ such that the total is summed in quadrature:

\begin{equation}
    3\sigma = \sqrt{(\sqrt{3\sigma})^2+(\sqrt{3\sigma})^2+(\sqrt{3\sigma})^2}
\end{equation}

From each source, we also remove data points that deviate 10$\sigma$ or greater from the mean as these outliers may be unphysical in origin (e.g. these outliers can appear when the reflection from solar panels comes into view). We demonstrate the effect of this  10$\sigma$ cutoff in \textit{Figure \ref{fig:10sigma}}. For some sources in the MAXI dataset, we found that very few to no data points remain once checked for 3$\sigma$ significance. This can happen because a source may vary in brightness and become fainter over time. For proper statistics, we considered only sources with at least 100 data points with 3$\sigma$ significance (\textit{Figure \ref{fig:100pt}}). 

Of the 58 sources from MAXI that we checked, only 12 BHs, 20 NPNSs, and 12 pulsars matched our criteria for statistical significance. These 44 sources are listed in \textit{Table \ref{table:1}}. The CCI diagram of all 44 sources are plotted in \textit{Figure \ref{fig:CCI}} and their corresponding 2D-projections are plotted in \textit{Figure \ref{fig:CCI_2D}}. \bedit{Since most of the BHs and some NPNS are transients, the sources that went into outburst during 16 years of RXTE/ASM may or may not go into outburst during the 10 years of MAXI operation. However, we have used new transients (which had not been active during the 16 years of RXTE/ASM), that have been discovered by MAXI.}


\subsection{Subsampling of Observations}
In our selection of data points, observations that are less bright are inherently less likely to be included. Since XRB systems of the same type can exhibit considerable variability in terms of which regions they occupy in CCI diagrams (\textit{Figure \ref{fig:CCI}}), this bias in data selection can be a limitation. We address this by subsampling where the probability that a particular observation is included in the training set is inversely proportional to the total number of observations of its system in the entire training set. An additional motivation for subsampling the data is for a computational reason: Gaussian process models can be particularly computationally expensive due to the inversion of a large matrix. The time required to perform this matrix inversion scales with $O (N^3)$ where N is the number of data points in the data set. 

\bedit{We run the subsampling algorithm 10 times to create 10 independent subsets of the data on which we can run our algorithms.} The histograms in \textit{Figure \ref{fig:subsamp}} show the number of observations between various systems before subsampling and five examples of the distribution after subsampling. For each independent run, we sample 20\% of the training data without replacement\bedit{, resulting in a dataset of 10,314 observations.}  

\section{Classification Algorithms}
In this paper, we compare a Bayesian Gaussian Processes \citep[BGP;][]{735807} method similar to the method used in GP15, K Nearest Neighbors \citep[KNN;][]{10.2307/2685209} and Support Vector Machines \citep[SVM;][]{Cortes95support-vectornetworks} in order to classify XRBs.  For each algorithm, we present the predictions for each compact object where the predicted class of an XRB is the class with the maximum estimated probability. We evaluate each algorithm in terms of its predictive accuracy for classifying XRBs in the MAXI/GSC dataset. Each algorithm is described in more detail in sections 3.1, 3.2, and 3.3.

\subsection{Bayesian Gaussian Process (BGP)}\label{bgp}
BGP methods have been extensively used in ML for classification and quantification of prediction uncertainty \citep{rasmussen2006cki}. In our work, we are using the BGP described in \citet{735807} as implemented in the R library Kernlab \citep{kernlab}.

To use a BGP to solve our classification problem, we must choose a covariance kernel. In ML, capturing the similarity between datapoints is essential; on a basic level, we assume that points with inputs $\mathbf{x}$ which are close are likely to have similar target $\mathbf{y}$ values. The covariance function determines how this proximity or similarity is defined, and in this paper we consider squared-exponential (i.e. Gaussian), Laplacian, and ANOVA radial basis function kernels as implemented in  \texttt{kernlab} \citep{kernlab}. All three kernels are variants of radial basis functions meaning that they are both \textit{stationary} and \textit{isotropic} kernels. \textit{Stationary} kernels are invariant to translations in the input space and are a function of $\mathbf{x -x'}$. If a kernel is also \textit{isotropic}, it is more specifically a function of $|\mathbf{x -x'}|$, making it invariant to rigid motions \citep{rasmussen2006cki}. We explicitly define the Gaussian and Laplacian Kernels below: 

\begin{equation}
    \textcolor{black}{K(\textbf{x},\textbf{x'}) = \text{exp}(- \frac{\mid\mid \textbf{x}-\textbf{x'}\mid\mid^2}{2\sigma^2}) \ \text{Gaussian}}
\end{equation}
\vspace{-0.1in}
\begin{equation}
    \textcolor{black}{K(\textbf{x},\textbf{x'}) = \text{exp}(- \frac{\mid\mid \textbf{x}-\textbf{x'}\mid\mid}{\sigma}) \ \text{Laplacian}}
\end{equation}

Where $\mid\mid \textbf{x}-\textbf{x'}\mid\mid$ refers to the Euclidean distance (i.e.\ the 2-norm) between two inputs $\textbf{x}$ and $\textbf{x'}$, $\sigma$ is the characteristic length scale. To see a full derivation of the ANOVA kernel, please refer to \citet{hofmann2008kernel}. In Section 4.2.2, we discuss how we compare and select from these three kernels for our specific classification problem. 

We determine the probability and estimate the errors in our BGP model by undertaking 5 independent runs of the best BGP model (\textit{Table \ref{table:A1}}) as described further in \textit{Section \ref{aggrege_preds}}. Although we did this for 10 runs for each of our other models, computational costs limited us to 5 runs for the BGP.  \textit{Table \ref{table:A1}} lists the median and standard deviation from the five runs. 

\begin{figure}
    \centering
    \epsscale{0.93}
    \plotone{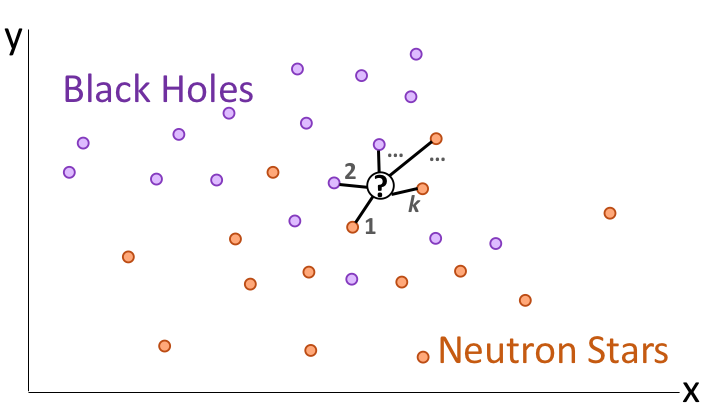}
    \caption{\footnotesize{A schematic depicting K-Nearest Neighbors (KNN) for BHs and NPNSs in two dimensions. For an unclassified source in the center, the model calculates the Euclidean distance between that point and every other data point. Then the k closest points to the unclassified source are used to predict the class. In this case, 60\% of the KNN are NPNSs while 40\% are BHs. Thus, the source has a 60\% probability of containing a NPNSs and 40\% probability of containing a BH. 
}
    \label{fig:knn_schem}}
\end{figure}

\begin{figure*}
\gridline{\fig{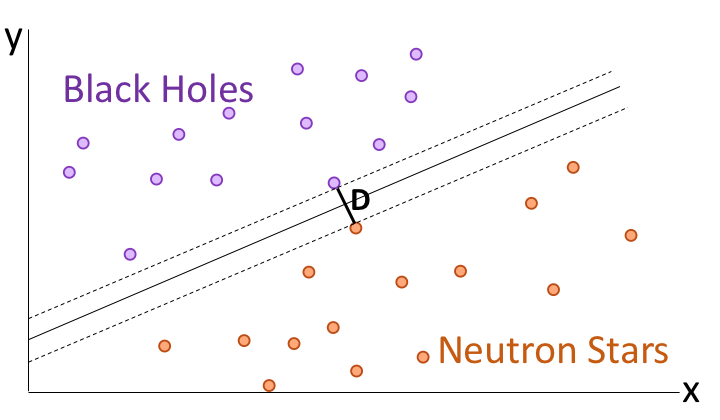}{0.33\textwidth}{(a)}
          \fig{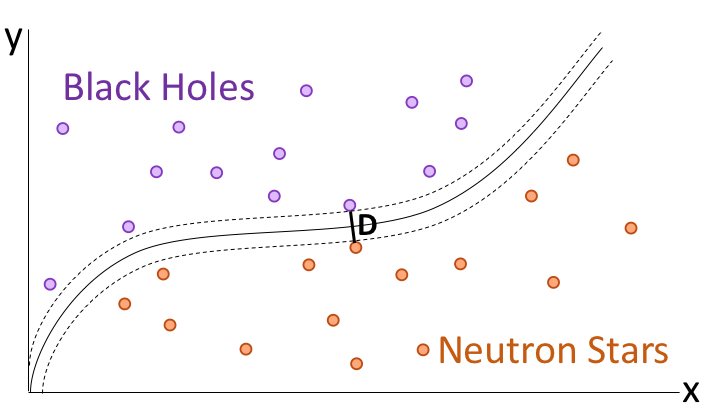}{0.3303\textwidth}{(b)}
          }
\caption{\footnotesize{Simple representations for Support Vector Machines (SVM). 6a illustrates how a linear SVM finds the line that creates the largest separation possible between two classes of data, BHs and NPNSs. 6b extends this to the nonlinear SVM which uses the ``kernel trick'' to project the points into more dimensions, resulting in a nonlinear curve. }
\label{fig:svm_schem}}
\end{figure*}

\subsection{K Nearest Neighbors (KNN)}

A simpler ML algorithm called KNN \citep{10.2307/2685209} is also used to predict XRB classifications. To implement our KNN algorithm, we customized code from the R class package \citep{classR}. For any given input CCI data point, the KNN model first calculates the Euclidean distance between that data point and every other point in the training set. Certain variables might have an unusually large impact on this distance because their magnitude is larger without normalization. For example, the relative intensity spans a larger range of values than soft colors and hard colors, meaning that any distances along the soft colors and hard colors could be overshadowed by distances along the relative intensity. To adjust for this, we normalized the dataset by the average of the top 0.01\% of the counts before we began. Normalizing the training data has been shown to improve the KNN algorithm's predictive accuracy dramatically \citep{doi:10.1061/JPEODX.0000175,Hast:Tibs:Frie:2001}.

After calculating the distances, we find the $k$ data points closest to the input point $p$, which are the $k$-nearest neighbors. Then, we classify the point $p$ based on the classifications of those neighbors. For example, if the majority of the $k$ closest data points are labelled  BH, then we classify the point as a BH. In \textit{Figure \ref{fig:knn_schem}}, KNN for a simple 2D case is illustrated where the two classes are BHs and NPNSs. Here, $k=5$ and for the unknown point $p$, 60\% of the $k$-Nearest Neighbors are NPNSs while 40\% are BHs. Thus, the source has a 60\% probability of being a NPNSs and 40\% probability to be a BH.

We choose $k$ with a cross-validation procedure described in Section \ref{crossval_descrip}. After choosing $k$, we determine the median probability of belonging to a certain class and estimate the errors in our KNN model by undertaking 10 independent runs.  Independence is achieved by generating a different subsample for each run (see \textit{Figure \ref{fig:subsamp}}).  \textit{Table \ref{table:A2}} lists the median and standard deviation from the 10 runs.


\subsection{Support Vector Machines (SVM)}

Finally, we compare the use of the SVM \citep{Cortes95support-vectornetworks} algorithm for making predictions of the compact object type of an XRB. To implement our SVM algorithm, we customized code from the LIBSVM library \citep{CC01a}. In a SVM, given the input and output data for each source, the goal is to find the hyperplane that creates the largest separation between the three classes of data. This is most clearly explained by examining a simple two-dimensional example. 

Consider \textit{Figure \ref{fig:svm_schem}a}, where data points are graphed with respect to variables x and y, and colored purple or orange depending on whether they belong to the BH or NPNSs class.
Given these inputs, a linear SVM model would calculate the two-dimensional line that
provides the largest possible separation between the two classes. In this case, the points border the two dotted lines and are a distance $D$ apart. The separator between the two classes is the solid line between them. New points are classified based on their location with respect to this line. Note that to create this model, only the points near the boundary determine the location of the separator. Thus, points far away from the boundary could be removed from the training set and the resulting model would be identical.

However, this method has the obvious problem that the points may not always be linearly
separable. To resolve this, a nonlinear SVM model uses a method known as the ``kernel trick'' where the points are projected into a space with even more dimensions, allowing the separator line to become a nonlinear curve. An example of this can be seen in \textit{Figure \ref{fig:svm_schem}b} where a nonlinear curve separates the BHs from NPNSs.

In higher dimensions, the logic is similar: the points are graphed in an n-dimensional
space, and the optimal (n-1)-dimensional dividing surface is then calculated. In the case of a linear SVM, this is a linear hyperplane; in the case of a nonlinear SVM, which we will use here, this is a nonlinear surface. 

We computed both the probability of belonging to a class and error estimates. SVMs can produce class probabilities as outputs by fitting a sigmoid function
\begin{equation}
    P(y=1 | f) = \frac{1}{1 + \exp{Af+B}}
\end{equation}
to the decision values $f$ of the binary SVM classifiers, where $A$ and $B$ are estimated by minimizing the negative log-likelihood function. For this multi-class case, all 3 binary classifier probability outputs can be combined as described in \citep{Wu}. This method is equivalent to fitting a logistic regression model to the estimated decision values \citep{Karatzoglou2006}.

 We determine the median probability of belonging to a certain class and estimate the errors in our SVM model by undertaking 10 independent runs.  Independence is achieved by generating a different subsample for each run (see \textit{Figure \ref{fig:subsamp}}).  \textit{Table \ref{table:A3}} lists the median and standard deviation from the 10 runs.

\section{Data  Analysis}
We use the sources from MAXI/GSC that meet our statistical significance criteria in section 2.1. We use a 3D representation of data as introduced by VB13 to test the three ML techniques described in section 3.

\subsection{Input Representation}
For each example in our training dataset, the input vector consists of the coordinates in the CCI diagram and the output vector is the class (1 = BH, 2 = NPNS, or 3 = pulsar). \bedit{The model is trained with these examples and asked to predict the probability that the input vector belongs to each of the classes. The output vector thus takes the form of  $Y_{pred} = (0.12, 0.73, 0.15)$ for an example that has a 12\%, 73\%, and 15\% probability of belonging to the BH, NPNS, and pulsar classes respectively. }

For the MAXI dataset, these examples can be described as 2-tuple ($X_{train}, Y_{train}$) where $X_{train}$ is the $N_{train}$ by 3 matrix and $Y_{train}$ is the classification corresponding to the compact object type 1, 2 or 3. Our inputs can thus be represented as 

\begin{equation}
\quad
\begin{bmatrix} 
SC_{1,1} & HC_{1,1} & Rel.Int_{1,1}\\
SC_{1,2} & HC_{1,2} & Rel.Int_{1,2} \\
SC_{1,3} & HC_{1,3} & Rel.Int_{1,3}  \\
... & ... & ...  \\
... & ... & ...  \\
... & ... & ...  \\

SC_1N_{1,train} & HC_1N_{1,train} & Rel.Int_1N_{1,train} \\
\end{bmatrix}
\quad
\end{equation}

where $SC_1$ stands for the soft color, $HC_1$ stands for the hard color, and $Rel.Int_1$ stands for the relative intensity of the first data point. The output $Y_{train}$ is a vector of length $N$ where $N$ refers to the number of statistically significant data points in the training set. \bedit{For a prediction for a single data point of an unclassified system, the ML algorithms use 3 features (SC, HC, Relint). Thus, for each row of $SC$, $HC$ and $Rel.Int$ values in Equation 8, the ML models output a probability distribution across the BH, NPNS, pulsar classes. All predictions belonging to the observations for a particular source are then aggregated to generate a median probability distribution per source as described in \textit{Section \ref{aggrege_preds}}. Thus, for a XRB source with $M$ observations, the number of features used to compute the overall probability distribution is $3 \times M$.}

To ensure each model generalizes to XRB examples it has never seen before, we use a cross validation method described in section 4.2. This method checks whether the model can be applied to predict out-of-sample data rather than only being capable of memorizing and reproducing the input training set.

\subsection{Cross-Validation and Model Hyperparameters}\label{crossval_descrip}

A fundamental challenge in ML is that algorithms must perform well on \textit{novel, previously unseen} inputs - not just examples on which the model was trained \citep{Goodfellow-et-al-2016}. Being able to perform well on \textit{novel, previously unseen} inputs is often referred to as generalization.

To estimate the ability for each ML algorithm to generalize to out-of-sample data, we use a cross validation procedure. At each iteration of the procedure, we remove the CCI points corresponding to a particular source and use the remaining sources as training data\bedit{, which comprises between 97 to 98 \% of the entire dataset depending on the number of observations belonging to the removed source}. Then, we predict the classification of the left-out source. We iterate this procedure for all of the sources in the data set, and the total number of correct classifications out of the 44 sources is indicative of the predictive accuracy on out-of-sample data. This cross validation procedure is similar to k-fold cross validation except that the data set is chunked based on the sources, as opposed to k equally sized chunks typically used in cross validation. Chunking the data by source is more realistic since the ML algorithms will ultimately be used to predict the classification of a new source that the algorithms were not trained on.  

\begin{deluxetable}{ll|ccc|c}[ht!]
\centering
\tabletypesize{\footnotesize}
\tablenum{2}
\tablecaption{BGP Cross-Validation Results \label{table:2}}
\tablewidth{1pt}
\tablehead{\multicolumn{2}{c|}{Model}  & \multicolumn{3}{c|}{Percentage Classified as X} & \multicolumn{1}{c}{Overall} \\ [0.5ex]
\multicolumn{2}{c|}{parameters} & \multicolumn{3}{c|}{Per Source Type X*} & \multicolumn{1}{c}{Accuracy}}
\startdata
\hline
Kernel & Time(hrs) & BH & NPNS & \multicolumn{1}{c|}{Pulsar} & Total \\
\hline
ANOVA  &  101.4 & 100.00\% & 5.00\%$^{\dagger}$  & 0.00\%$^{\dagger}$ & 29.55\%    \\
Laplace & 32.7  & 41.67\%  & 95.0 \% & 100.00\% & 81.82\%    \\
Gaussian &  132.6  & 41.67\%  & 95.0 \% & 100.00\% & 81.82\%    \\
\enddata
\tablecomments{\footnotesize{We determined the model accuracy across different kernels, by  implementing each one and counting the number of correct classifications for each source type.}}
\tablenotetext{*}{\footnotesize{Here X can be a BH, NPNS, or pulsar. For example, for the BH column,  the percentage listed is the percentage of BH sources that are classified as BHs. For the NPNS column, the percentage listed is the percentage of NPNS sources that are classified as NPNSs, etc.}}
\tablenotetext{\dagger}{\footnotesize{Only 5\% of NPNSs are classified as NPNSs and 0\% of pulsars are classified as pulsars because the algorithm classified nearly all sources as BHs.
\vspace{-0.3in}}}
\end{deluxetable}

\begin{deluxetable}{lc|ccc|c}[h!]
\centering
\tabletypesize{\footnotesize}
\tablenum{3}
\tablecaption{KNN Cross-Validation Results \label{table:3}}
\tablewidth{1pt}
\tablehead{\multicolumn{2}{c|}{Model}  & \multicolumn{3}{c|}{Percentage Classified as X} & \multicolumn{1}{c}{Overall} \\ [0.5ex]
\multicolumn{2}{c|}{parameters} & \multicolumn{3}{c|}{Per Source Type X*} & \multicolumn{1}{c}{Accuracy}}
\startdata
\hline
$K$ values & Time (mins) & BH & NPNS & \multicolumn{1}{c|}{Pulsar} & Total \\
\hline
2 - 16 & $\sim$ 10 & 41.67\%$^\dagger$ & 90.00\% & 100.00\% & 79.54\% \\
17 - 18 & $\sim$ 10 & 50.0\% & 90.00\% & 100.00\% & 81.82\% \\
19 - 21 & $\sim$ 10 & 41.67\% & 90.00\% & 100.00\% & 79.54\% \\
22 & $\sim$ 10 & 50.0\% & 90.00\% & 100.00\% & 81.82\% \\
23 & $\sim$ 10 & 41.67\% & 90.00\% & 100.00\% & 79.54\% \\
24 & $\sim$ 10 & 50.0\% & 90.00\% & 100.00\% & 81.82\% \\
25 - 43 & $\sim$ 10 & 41.67\% & 90.00\% & 100.00\% & 79.54\% \\
44 & $\sim$ 10 &  41.67\% & 95.00\% & 100.00\% & 81.82\% \\
45 & $\sim$ 10 & 41.67\% & 85.71\% & 100.00\% & 79.54\% \\
45-50 & $\sim$ 10 &  41.67\% & 95.00\% & 100.00\% & 81.82\% 
\enddata
\tablecomments{\footnotesize{We determined the model accuracy across values of k, by iteratively trying values from 2 to 50 and counting the number of correct classifications per source class.}}
\tablenotetext{*}{\footnotesize{Here X can be a BH, NPNSs, or pulsar. For example, for the BH column,  the percentage listed is the percentage of BH sources that are classified as BHs. For the NPNSs column, the percentage listed is the percentage of NPNSs sources that are classified as NPNSs, etc.}}
\tablenotetext{\dagger}{\footnotesize{For the BHs that are not correctly classified, the algorithm commonly predicts these sources to be NPNSs instead.
\vspace{-0.3in}}}
\end{deluxetable}

\subsubsection{BGP parameter tuning}

During this cross-validation process we  tune the hyperparameters of the various ML architectures. Each architecture has different hyperparameters that can be tuned, which potentially decrease generalization error and thereby improve model performance. Some of these hyperparameters are specifically designed to minimize generalization error and prevent the ML model from overfitting to the training data. Methods that prevent overfitting are called regularization methods and we will use cross-validation to optimize some of these regularization parameters.

For the BGP cross-validation, we compared the performance of three different covariance kernels (Gaussian, Laplacian, and ANOVA radial basis function kernels). The three kernels are explicitly defined in Section \ref{bgp}.

\begin{deluxetable*}{lc|c|ccc|cc}[ht!]
\centering
\tabletypesize{\footnotesize}
\tablenum{4}
\tablecaption{SVM Cross-Validation Results \label{table:4}}
\tablewidth{1pt}
\tablehead{\multicolumn{3}{c|}{Model}  & \multicolumn{3}{c|}{Percentage Classified as X} & \multicolumn{1}{c}{Overall} \\ [0.5ex]
\multicolumn{3}{c|}{Parameters} & \multicolumn{3}{c|}{Per Source Type X*} & \multicolumn{1}{c}{Accuracy}}
\startdata
\hline
C   & Gamma & Time (mins) & BH & NPNS & \multicolumn{1}{c|}{Pulsar} & Total \\
\hline
0.040 & 0.484 & $\sim$ 20 & 50.00\% & 95.00\% & 100.00\%    & 84.09\%    \\
0.116 & 0.318 & $\sim$ 20 & 50.00\% & 95.00\% & 100.00\%    & 84.09\%    \\
0.184 & 0.406 & $\sim$ 20 & 50.00\% & 95.00\% & 100.00\%    & 84.09\%    \\
0.231 & 6.437 & $\sim$ 20 & 50.00\% & 95.00\% & 100.00\%    & 84.09\%    \\
0.371 & 0.270 & $\sim$ 20 & 50.00\% & 95.00\% & 100.00\%    & 84.09\%    \\
0.397 & 1.898 & $\sim$ 20 & 50.00\% & 95.00\% & 100.00\%    & 84.09\%    \\
0.870 & 2.859 & $\sim$ 20 & 50.00\% & 95.00\% & 100.00\%    & 84.09\%    \\
1.969 & 0.051 & $\sim$ 20 & 50.00\% & 95.00\% & 100.00\%    & 84.09\%    \\
2.162 & 0.126 & $\sim$ 20 & 50.00\% & 95.00\% & 100.00\%    & 84.09\%    \\
0.020 & 5.141 & $\sim$ 20 & 58.33\% & 90.00\% & 100.00\%    & 84.09\%    \\
0.021 & 7.729 & $\sim$ 20 & 58.33\% & 90.00\% & 100.00\%    & 84.09\%    \\
\hline
0.062 & 2.314 & $\sim$ 20  & 58.33\% & 95.00\% & 100.00\%    & 86.36\%    \\
0.093 & 1.572 & $\sim$ 20 & 58.33\% & 95.00\% & 100.00\%    & 86.36\%    \\
0.100 & 0.579 & $\sim$ 20 & 58.33\% & 95.00\% & 100.00\%    & 86.36\%    \\
0.655 & 0.585 & $\sim$ 20 & 58.33\% & 95.00\% & 100.00\%    & 86.36\%  
\enddata
\tablecomments{\footnotesize{To compute how the model accuracy changed based of values of the hyperparameters C and Gamma, we implemented latin hypercube sampling across the parameter space (C = $2^{-15}$ to $2^{5}$, gamma = $2^{-15}$ to $2^3$). The 15 best performing combinations of C and gamma are summarized here.}}
\tablenotetext{*}{\footnotesize{Here X can be a BH, NPNS, or pulsar. For example, for the BH column,  the percentage listed is the percentage of BH sources that are classified as BHs. For the NPNS column, the percentage listed is the percentage of NPNS sources that are classified as NPNSs, etc.
\vspace{-0.3in}}}
\end{deluxetable*}

For all three kernels, the adjustable parameter $\sigma$ significantly affects the performance of the kernel and serves as a regularization parameter. We set the BGP to use automatic sigma estimation as implemented in \citep{kernlab}. Although the Gaussian and Laplacian kernels are closely related, the Laplacian kernel loses the square norm and is less sensitive to changes in $\sigma$. The ANOVA kernel has been shown to work particularly well for multidimensional regression problems \citep{dc5dea99379f4d019b0655de6d6385e3} which we hoped might be promising for this multidimensional classification problem.

As illustrated in \textit{Table \ref{table:2}}, the Gaussian and Laplacian kernels have the highest predictive accuracy with 81.8$\%$ of sources correctly classified whereas the ANOVA kernel only predicted 29.6$\%$ of sources correctly. The ANOVA predicted nearly every source to belong to the BH class which negates the high predictive accuracy of BH sources compared to NPNSs and pulsars. Although the Laplacian and Gaussian kernels provide the same number of correct predictions (81.8$\%$), the Gaussian kernel takes four times as long to run. Thus, we chose the Laplacian kernel for our final model since it is both the most computationally efficient and provides the best performance on the cross-validation. We ran our final model 5 times and computed the median and SD across those five runs (\textit{Table \ref{table:A1}}).

\subsubsection{KNN hyperparameters tuning}

For the cross-validation procedure, we chose to iterate through 49 different values for k. We examined the range $k=2$ through $k=50$ as these values are not so small that it leads to unstable decision boundaries, but also not so large that it becomes computationally expensive.

As illustrated in \textit{Table \ref{table:3}}, we found that for values of $k=17, 18, 22, 24, 44, 45-50$, 81.8$\%$ of sources are correctly classified. For these values of $k$, We found that $k =24$ provides predictions with the highest probabilities of belonging to the correct class across all three source types. Thus, we chose $k =24$ for our final model. We ran our final model 10 times and computed the median and SD across those ten runs (\textit{Table \ref{table:A2}}).

\begin{figure}[h!]
\centering
\includegraphics[width=0.75\linewidth]{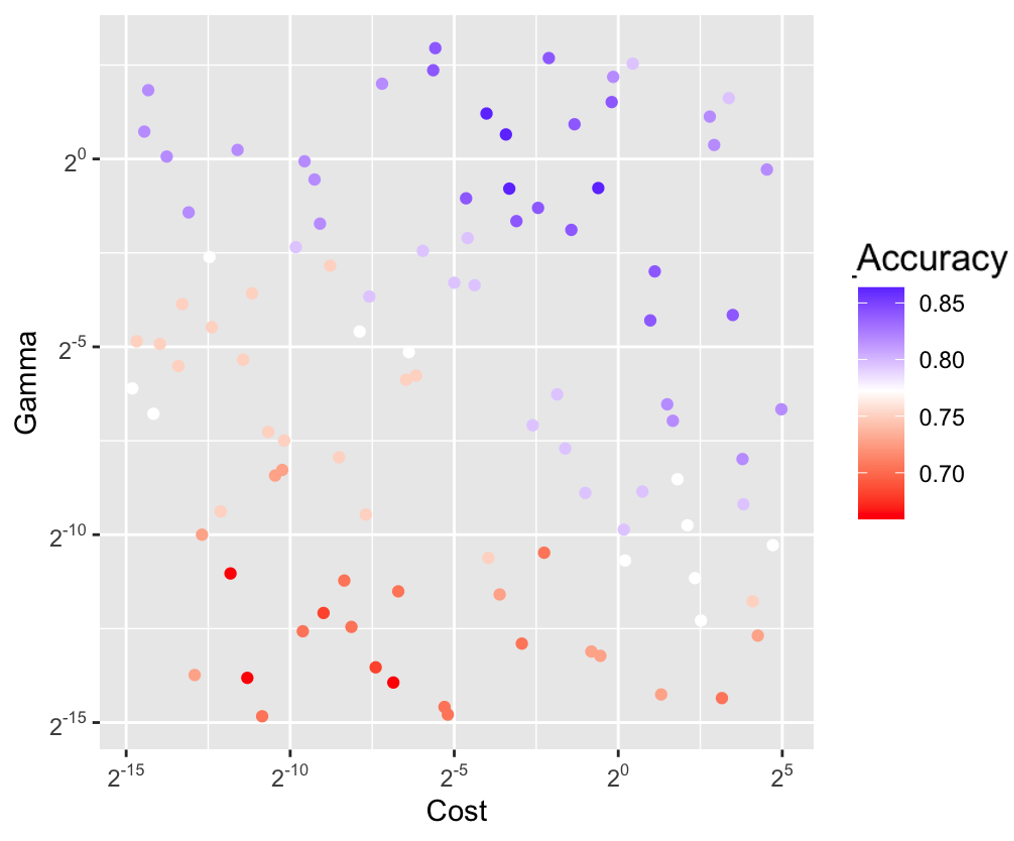}
\caption{\footnotesize{Latin hypercube sampling across values of Gamma and C. We ran 100 different configurations of Gamma and C for our SVM model using latin hypercube sampling. As both gamma and C increase, the number of correct classifications across sources (Accuracy) increases up until a certain point (Gamma $\sim 2^1$, Cost $\sim 2^{-3}$).}
\label{fig:hypercube}}
\end{figure}

\begin{figure*}[h!tp]
\centering
\includegraphics[height=0.35\textwidth]{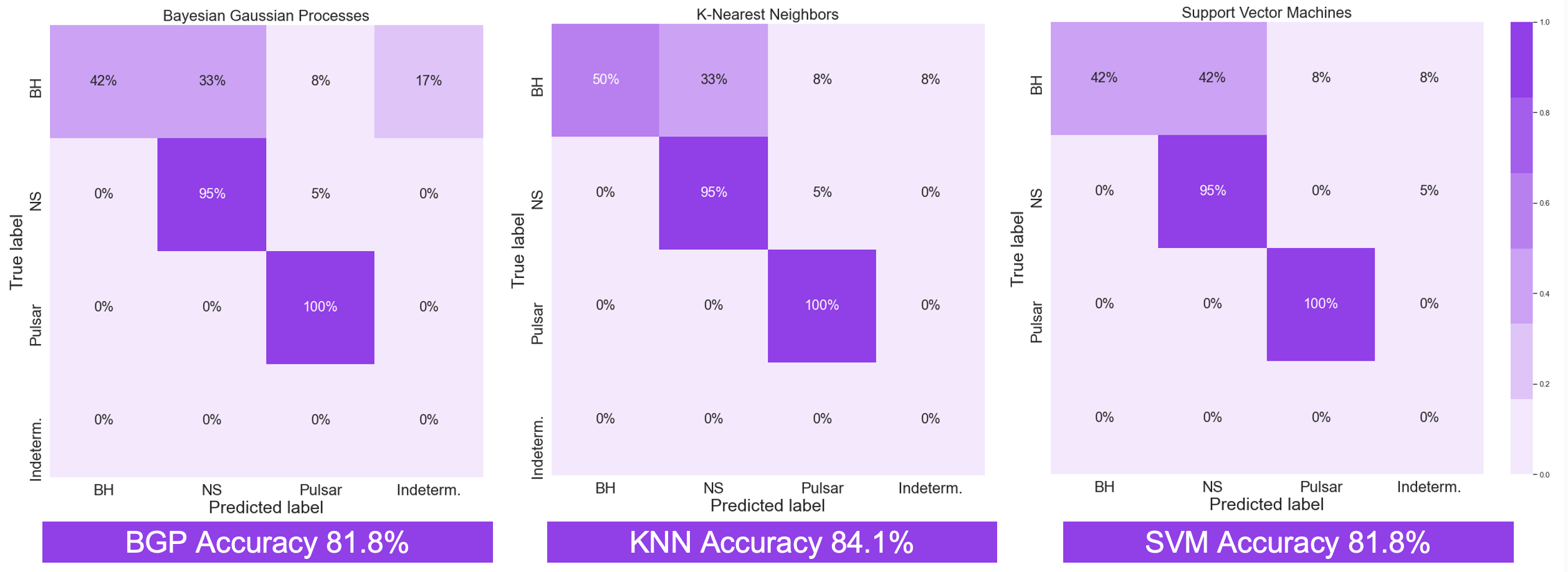}
\caption{\footnotesize{Accuracy of predictions per class of XRBs for each ML model. ``Indeterm.'' stands for indeterminate classifications where there is an equal probability (within error) of belonging to two classes. The KNN model has the highest predictive accuracy by predicting 84.1 $\%$ of sources in the cross-validation correctly. The BGP and SVM follow with 81.8$\%$ of cross-validation sources predicted correctly} (\textit{Figures \ref{fig:bhs_probs}, \ref{fig:ns_probs}, \ref{fig:pulsars_probs}}). All three models incorrectly classify a large fraction ($\sim$50 to 58 \%) of the BH sources.
\label{fig:summary}}
\end{figure*}

\subsubsection{SVM hyperparameters tuning}

For the SVM cross-validation, we chose to sample 100 values across the parameter space of the hyperparameters gamma and C. Since these parameters span a large range of possible ideal values ($2^{-15}$ to $2^3$\ for gamma; \citealt{CC01a}; $2^{-15}$ to $2^{5}$ for C; \citealt{CC01a} for C), they required an efficient sampling method. We implemented latin hypercube sampling \citep{lhs_package}, which is a statistical technique that achieves near-random sampling across parameters \citep{stein1987large} and is commonly used for sampling large parameter spaces.

Gamma can be thought of as defining the radius of influence of the support vectors. When gamma is very small, the model is too constrained and cannot capture the features or complexity in the data. However, making gamma too large corresponds to the radius of the support vectors only including the support vectors themselves, resulting in overfitting on the training data and large generalization errors.

The other parameter we simultaneously optimized was C, which is a regularization term in the Lagrange formulation of SVMs. C presents a trade-off between correct classification of training examples and against maximizing the margin of the decision function. Increasing C means that a smaller margin will be deemed acceptable and result in better classification of training points. Lower values of C will create a larger margin and a simpler decision function which may prevent overfitting, but also decreases training accuracy. In this way, C can prevent the SVM model from overfitting, decrease the generalization error and serve as a regularization parameter. 

The best performing hyperparameters combinations are summarized in \textit{Table \ref{table:4}} and the performance across all 100 samples is illustrated in \textit{Figure \ref{fig:hypercube}}. Ultimately, the best performing models with $\sim 86.4\%$ accuracy had values of C between $\sim$0.06 and $\sim$0.1 and for gamma between $\sim$0.57 and $\sim$2.3. 
For these values of gamma and C, We found that C = 0.655 and gamma = 0.585 provides predictions with the highest probabilities of belonging to the correct class across all three source types. Thus, we chose those hyperparameter values for our final model. We ran our final model 10 times and computed the median and SD across those ten runs (\textit{Table \ref{table:A3}}).

\subsection{Aggregation of Predictions}\label{aggrege_preds}
Since the ML algorithms treat each observation independently and compute their predicted probability distribution, we aggregate the predictions per source to compute an overall probability distribution. Specifically, for given XRB system with $M$ observations, we compute the overall probability distribution and corresponding standard deviations after performing the cross-validation by taking the following two steps:
\begin{enumerate}
    \item First, we compute the mean probability distribution across the $M$ observations belonging to that system. 
    \item  We then estimate the errors in our models by undertaking 10 independent runs (five for the BGP due to computational constraints). Independence is achieved by generating a different subsample for each run (see \textit{Figure \ref{fig:subsamp}}).  Across those ten independent runs, we repeat step 1, aggregate those results, and compute the median and sample standard deviation across the 10 probability distributions.
\end{enumerate}
Resampling data to estimate errors has a long history in statistics through methods such as bootstrapping \citep{Efron1993-mx} and has also been shown to reliably produce error estimates when applied to machine learning algorithms \citep{Sandia}.

\bedit{In this way, we aggregate the predictions to get a median and standard deviation for each XRB source in \textit{Table \ref{table:1}}.}

\subsection{\bedit{Indeterminate Classifications}}
\bedit{For some of the XRB sources, the median probability of belonging to two or more classes is equal (within 1$\sigma$). In this case, we chose to assign the sources an indeterminate classification since the algorithms cannot distinguish them within error. }

\section{Results of ML Classification}
\begin{figure*}[h!tp]
\centering
\includegraphics[clip,height=0.85\textwidth]{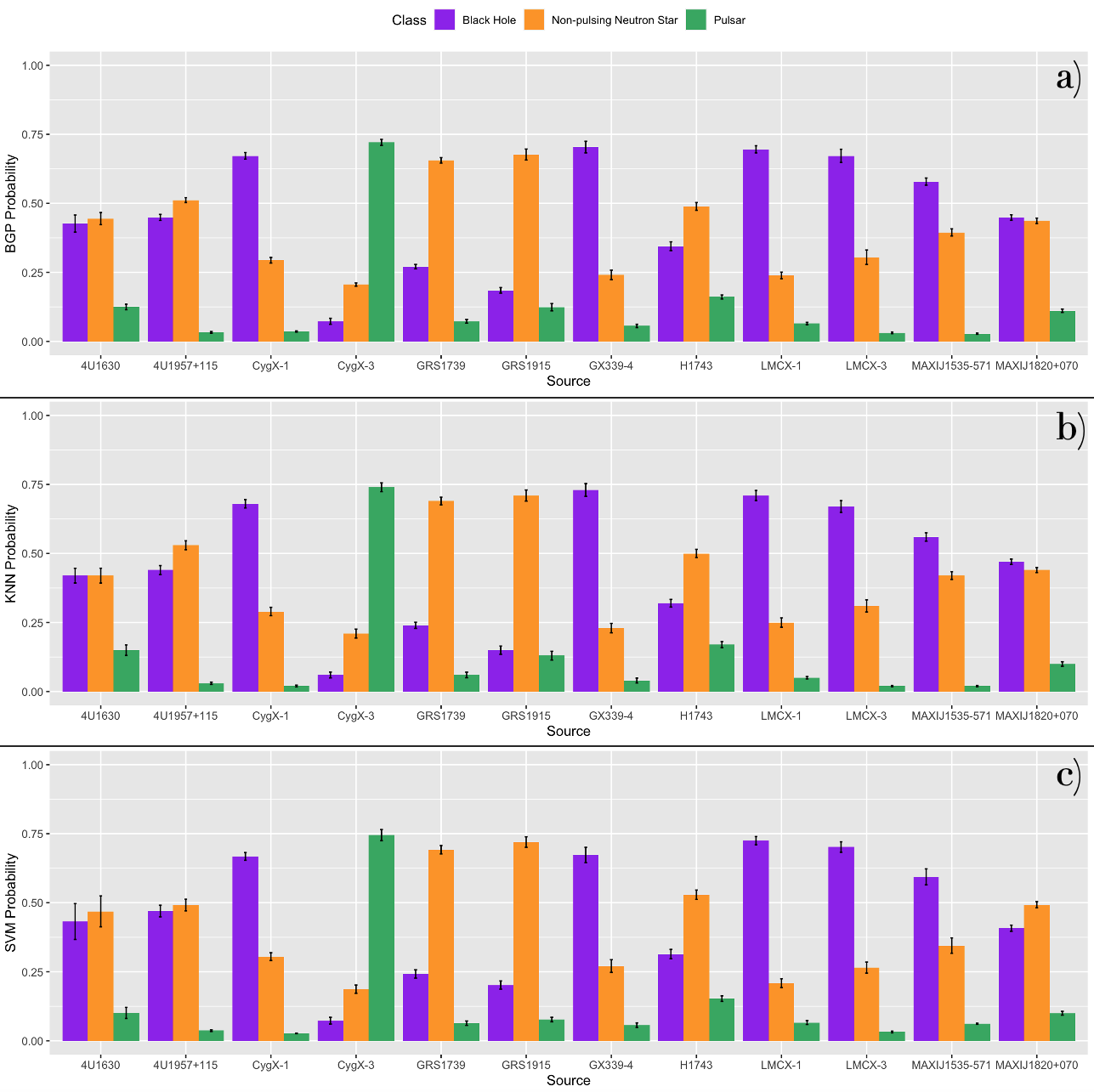}
\caption{\footnotesize{Probabilities of 44-source cross-validation for BH sources derived from ML models. Purple indicates the probability of belonging to the class of BHs, orange indicates NPNSs, and green indicates pulsars. The median probability across ten runs and the standard deviation across runs are plotted.}
\label{fig:bhs_probs}} 
\end{figure*}

\begin{figure*}[h!tp]
\centering
\includegraphics[clip, height=0.85\textwidth]{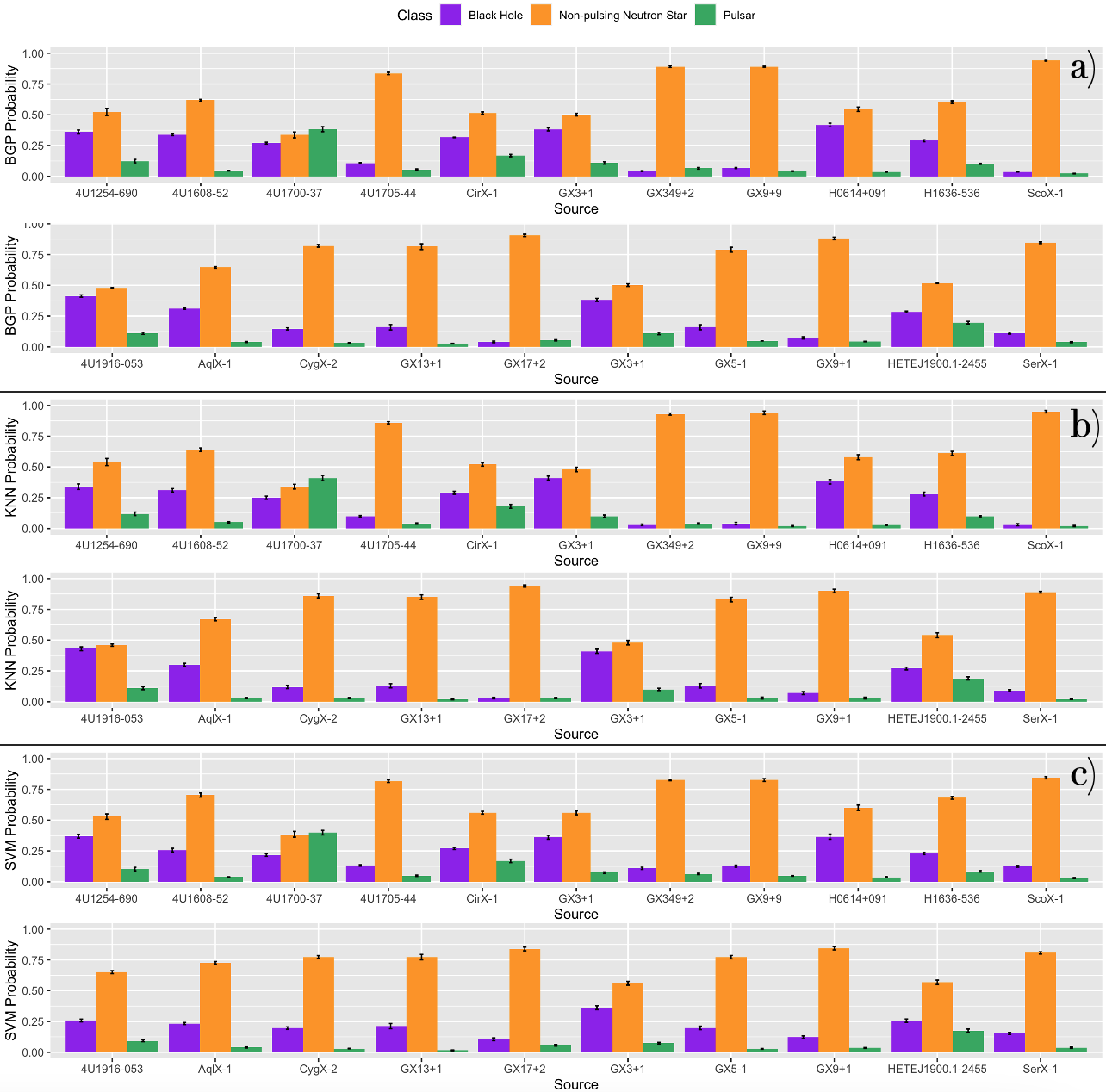}
\caption{\footnotesize{The probabilities of 44-source cross-validation for NPNS sources derived from all three ML models. Purple indicates the probability of belonging to the class of BHs, orange indicates NPNSs, and green indicates pulsars. The median probability across ten runs and the standard deviation across runs are plotted.}}
\label{fig:ns_probs}
\end{figure*}

\begin{figure*}[h!tp]
\centering
\includegraphics[clip, height=0.85\textwidth] {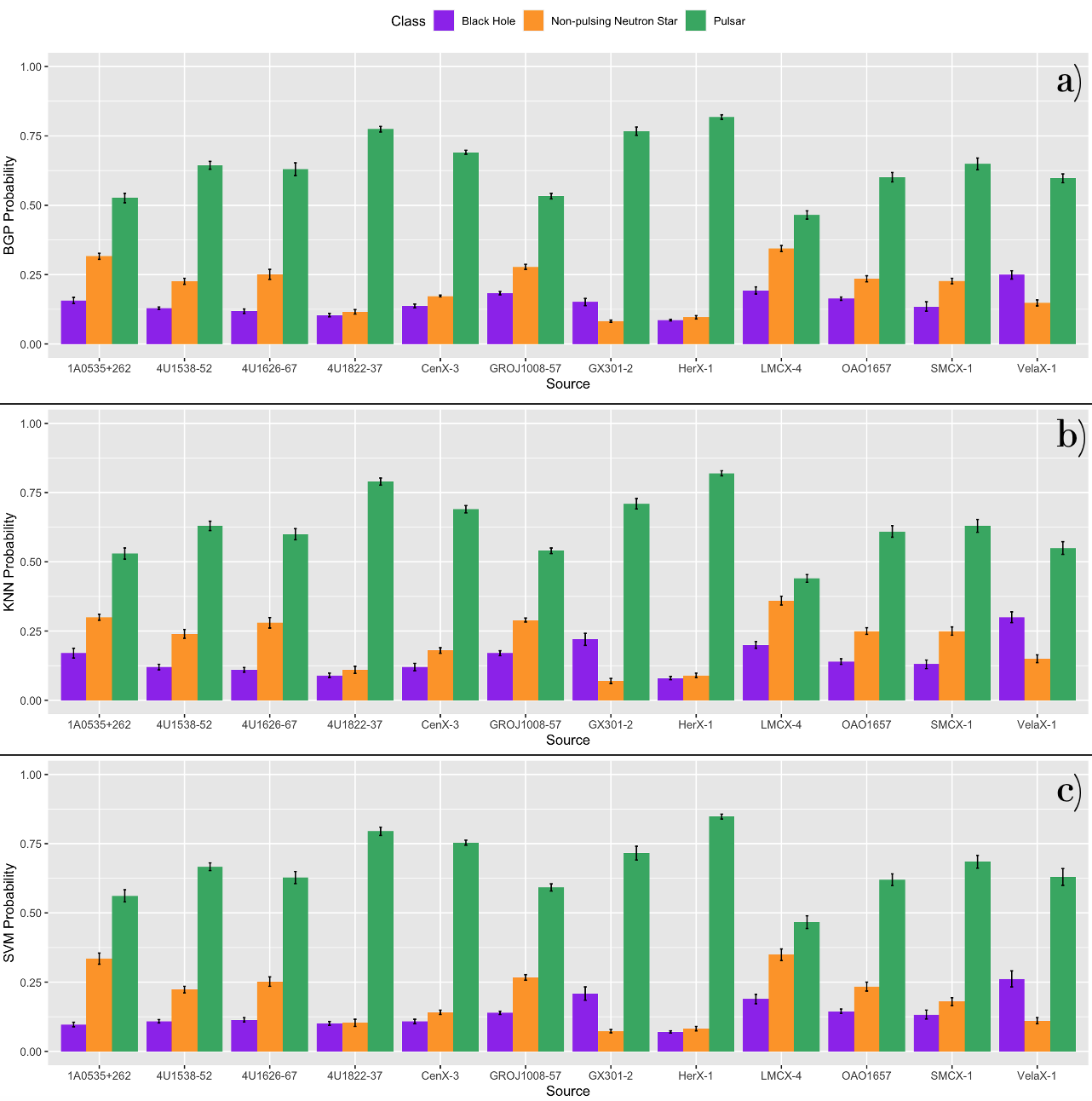}
\caption{\footnotesize{The probabilities of 44-source cross-validation for pulsar sources derived from all three ML models. Purple indicates the probability of belonging to the class of BHs, orange indicates NPNSs, and green indicates pulsars. The median probability across ten runs and the standard deviation across runs are plotted.}}
\label{fig:pulsars_probs}
\end{figure*}

In \textit{Figure \ref{fig:summary}}, we summarize the predictive accuracy for each algorithm and XRB class \bedit{in confusion matrices, which were generated from the average predictions across all runs\footnote{See \textit{Section \ref{aggrege_preds}}  on how these average predictions were computed.}}. All three methods have a relatively high predictive accuracy on average, but vary considerably across XRB classes. The KNN performs best with 84.1$\%$ of sources correctly classified compared to 81.8\% for both the BGP and the SVM. \textit{Figures \ref{fig:bhs_probs}-\ref{fig:pulsars_probs}} depict the probabilities from the 44-source cross-validation for XRB sources derived from the three ML models. Purple indicates the probability of belonging to the class of BHs, orange indicates NPNSs, and green indicates pulsars. All three methods incorrectly classify a large fraction ($\sim$50 to 58 $\%$) of the BH sources.


Based on these results, we can use our prior knowledge to compute the conditional probabilities that a source is a specific type of compact object given the prediction (often also referred to as precision). For example, we can compute the probability that a source's true classification, $true_{class}$, is a NPNS given that our BGP model predicts it to be a NPNS in the following way:

\begin{center}
\begin{multline}
     P(true_{class} = NPNS\ {\vert pred_{class}  = NPNS)} =  \\
     \frac{\vert true_{class} = NPNS, pred_{class} = NPNS \vert}{\vert pred_{class} = NPNS \vert} = \frac{19}{23} = 0.83\\
\end{multline}
\end{center}

where $pred_{class}$ is the model predicted classification, 23 of our sources are predicted to be NPNSs and only 19 both have a true classification and predicted classification as a NPNS. Thus, the BGP conditional probability that a source predicted to be a NPNS truly is a NPNS is 0.83. In the exact same way, we computed these BGP model conditional probabilities for BHs and pulsars and for all three classes using KNN and SVM as presented in \textit{Table \ref{table:5}}.

\begin{deluxetable}{l|cc}[ht!]
\centering
\tablenum{5}
\tablecaption{Conditional Probabilities \label{table:5}}
\tablewidth{1pt}
\tablehead{\multicolumn{1}{c|}{BGP Conditional}  & \multicolumn{1}{c}{Probability}}
\startdata
\hline
$P(true_{class} = BH\ {\vert\ pred_{class}  = BH)}$  & 100.0\% \\
$P(true_{class} = NPNS\ {\vert\ pred_{class}  = NPNS)}$  & 82.6\% \\
$P(true_{class} = Pulsar\ {\vert\ pred_{class}  = Pulsar)}$ & 85.7\% \\[0.5ex]
\hline
\multicolumn{1}{c|}{KNN Conditional} & Probability \\
\hline
$P(true_{class} = BH\ {\vert\ pred_{class}  = BH)}$  & 100.0\% \\
$P(true_{class} = NPNS\ {\vert\ pred_{class}  = NPNS)}$  & 82.6\% \\
$P(true_{class} = Pulsar\ {\vert\ pred_{class}  = Pulsar)}$ & 85.7\% \\[0.5ex]
\hline
\multicolumn{1}{c|}{SVM Conditional} & Probability \\
\hline
$P(true_{class} = BH\ {\vert\ pred_{class}  = BH)}$  & 100.0\% \\
$P(true_{class} = NPNS\ {\vert\ pred_{class}  = NPNS)}$  & 79.2\% \\
$P(true_{class} = Pulsar\ {\vert\ pred_{class}  = Pulsar)}$ & 92.3\% \\
\enddata
\tablecomments{We determined the conditional probabilities for each model and source type using \textit{Equation 9}.}
\end{deluxetable}

\begin{figure}
\centering
\includegraphics[width=0.9\linewidth]{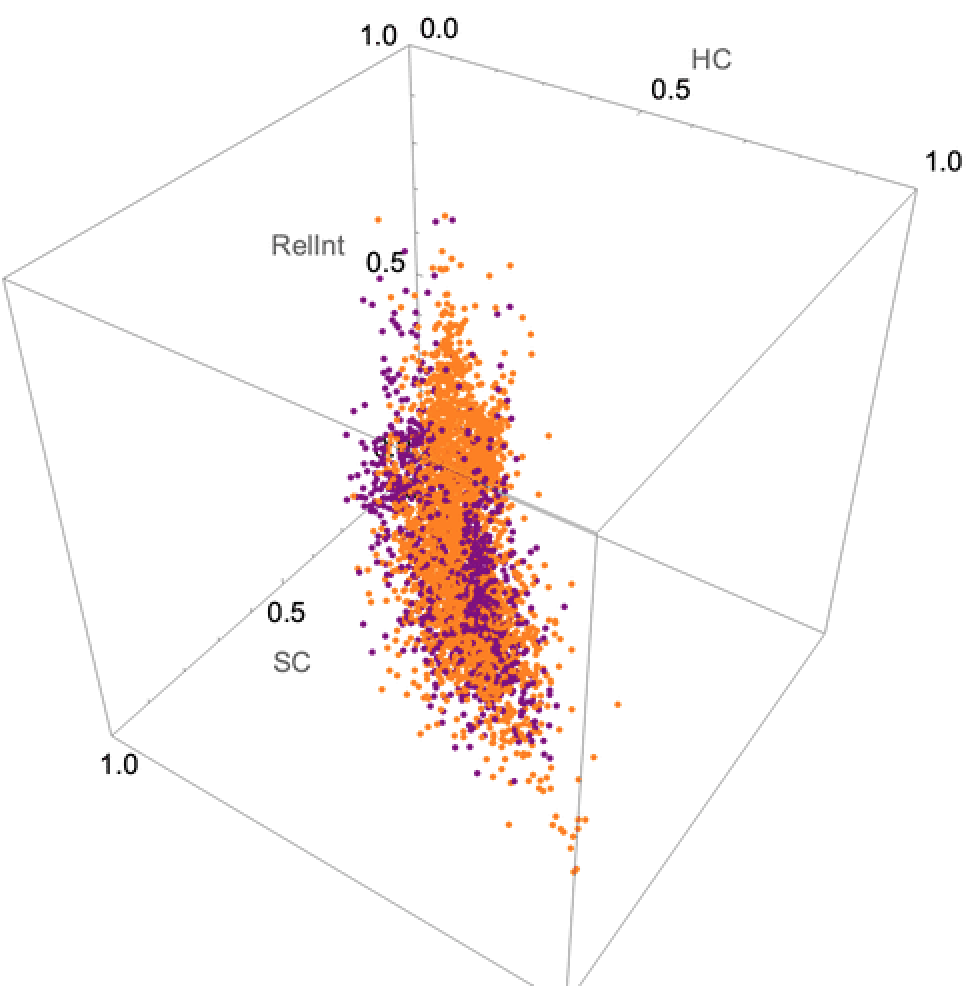}
\caption{The four BHs misclassified as Bursters in purple and Bursters in orange.
\label{fig:bhs_as_bursters}}
\end{figure}

\begin{figure}
\centering
\includegraphics[width=0.9\linewidth]{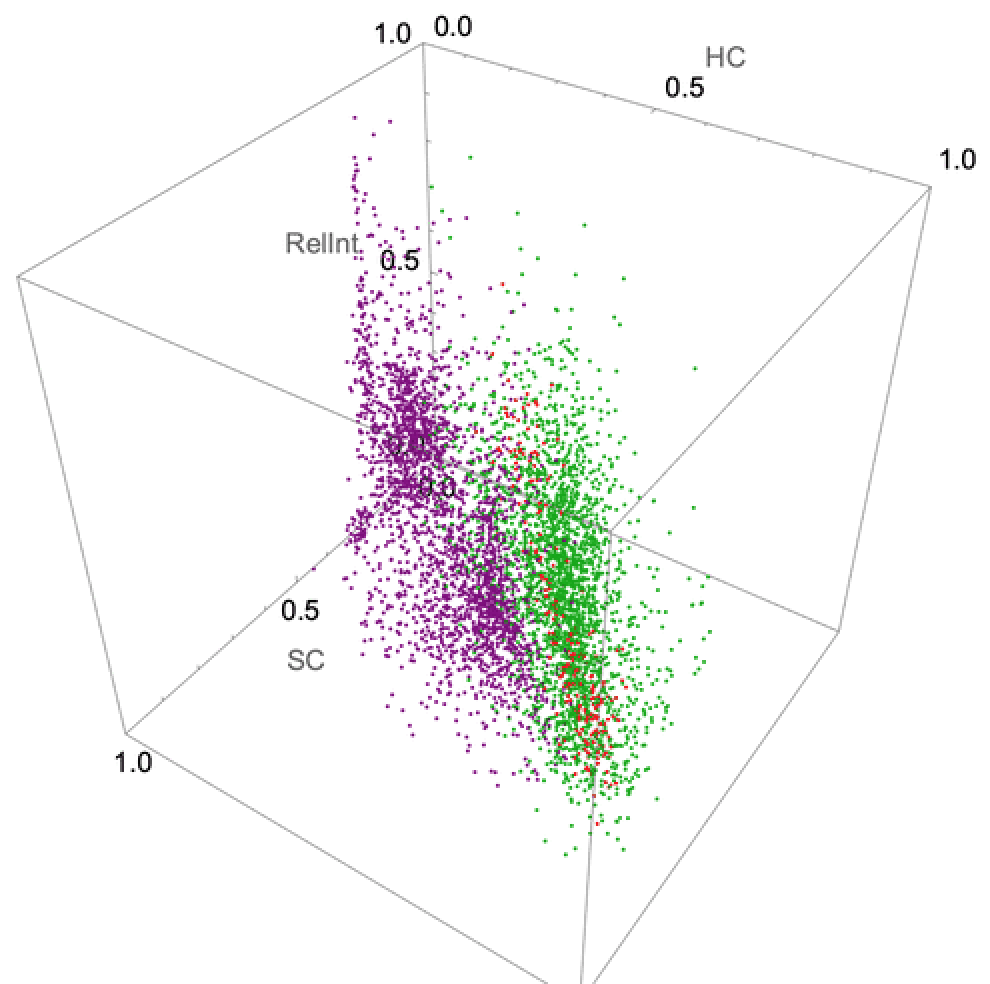}
\caption{ The BH Cyg X-3 in red; BHs in training set excluding Cyg X-3 in purple; all pulsars in training set in green.
\label{fig:cygx3}}
\end{figure}

\begin{figure}
\centering
\includegraphics[width=0.9\linewidth]{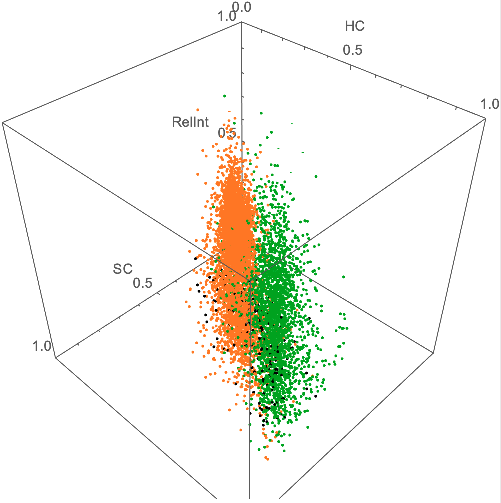}
\caption{The NPNS 4U1700-37 in black, other NPNS sources in the training set in orange, and all pulsar sources in training set in green.
\label{fig:4u1700}}
\end{figure}

In all three methods the primary misclassification of BHs are as NPNSs.  The difficulty of separating BHs from a subclass of NPNS called Bursters was already noted by GP15.  As \textit{Figure \ref{fig:10sigma}} illustrates both BHs and Bursters spend significant time in quiescence.  In \textit{Figure \ref{fig:bhs_as_bursters}}, we plot the four BHs that are misclassified as NPNSs together with the Bursters.  It is clear that Bursters and BHs are distinct in location during their ``on'' or ``burst'' states, while there is significant overlap during quiescence.  One possibility to mitigate this effect is to put in an intensity cutoff for Bursters and/or BHs in the training set. We tested this approach for Bursters and describe the results in \textit{Section \ref{bh_misclass_cutoff}}.  A more universal means of mitigating this effect would be to include a fourth dimension that takes into account the temporal variability of the sources. From the examples in \textit{Figure \ref{fig:10sigma}} it appears that BHs spend more time in quiescence with ``on'' states that last longer than the durations of the ``bursts'' of Bursters.  However, a more rigorous definition of variability may be necessary.  Both these methods are currently being tested and will be presented in a future paper.

All three models misclassify Cyg X-3 as a pulsar. In \textit{Figure \ref{fig:cygx3}} we show Cyg X-3 in red, along with BHs (excluding Cyg X-3) in purple, and all pulsars in green. It is clear that Cyg X-3 has little overlap with the other black holes in our sample but does have some overlap with pulsars.  Cyg X-3 is an unusual source, showing ultra-high energy gamma-rays and radio flares during which it becomes the brightest radio source in the Milky Way \citep{2018A&A...612A..27K}.  

All three models do well on classifying NPNS with one exception: 4U1700-37 is classified as a pulsar (or as indeterminate\footnote{The SVM predicts that there is equal probability (within error) that 4U1700-37 is a NPNS or Pulsar.} by the SVM).  In \textit{Figure \ref{fig:4u1700}}, we show 4U1700-37 in black, along with all NPNS in orange, and all pulsars in green. It is clear that while 4U1700-37 does share some space with NPNS, the majority of its data points overlap with pulsars. While no pulses have been detected from 4U1700-37, several authors have noted that its temporal and spectral characteristics are very similar to highly magnetized HMXB \citep{2018MNRAS.473L..74M, 2016MNRAS.461..816I, 2003ApJ...592..516B}. 

All three models have no difficulty in correctly classifying all pulsars (\textit{Figure \ref{fig:pulsars_probs}}).  

\subsection{BHs misclassification as NPNSs: Quiescence analysis}\label{bh_misclass_cutoff}

To examine the role of the quiescent observations in the misclassification of BHs as NPNSs, we ran two independent analyses where we included only the Burster observations when sources were in a ``on" or ``burst" state and all observations for other source types. In these two analyses, we changed how we determined which points were considered in quiescence versus in a ``burst" state.

In the first analysis, we applied a uniform cutoff where all points with a relative intensity below 0.4 were removed from Burster sources in the training set. The number 0.4 was chosen since most ``on" or ``burst" observations fall above this threshold. We found that this uniform cutoff significantly increased the BH classification accuracy, leading to twice as many BHs being classified correctly for all three ML methods. In addition, pulsars and NPNSs are correctly classified with rates of 100\% and 90\% respectively.

In the second analysis, we changed the exact cutoff based on each Burster source's unique variation in relative intensity. In particular, we computed the mean and standard deviation ($SD$) of the relative intensity for each Burster source. Then, we define the cutoff as,

\begin{equation}
    \textcolor{new_color}{\text{cutoff}_{2\sigma} = mean (\text{RelInt}) + 2\cdot SD(\text{RelInt})}
\end{equation}
where $RelInt$ is the relative intensity. We assume all points below this cutoff to be in quiescence and remove them from the training set. Similar to the 0.4 cutoff analysis, we find that this $\text{cutoff}_{2\sigma}$ method also  leads to twice as many BHs being classified correctly for all three ML methods. Pulsars and NPNSs are correctly classified with rates of 100\% and 85\% respectively. Overall, the results across these two quiescense analyses supports our hypothesis that significant parameter overlap during quiescence of BHs and NPNS could play a role in the misclassifications we observe for BHs.

\subsection{Probability Thresholds}
The accuracy of each of the ML algorithms can also be affected by the choice of a specific probability threshold. In the results reported throughout this paper, the class with the largest probability is assigned as the predicted class, regardless of the exact value of the largest probability. However, in some machine learning applications, a probability threshold is chosen instead to assign the predicted class. We investigated how choosing a probability threshold rather than choosing the class with the largest predicted probability would affect the accuracy of our models by using Receiver Operator Characteristic (ROC) curves.

ROC curves plot the True Positive Rate (TPR) and False Positive Rate (FPR) for various probability thresholds. The TPR and FPR are defined in the following way:
\begin{equation}
    \textcolor{new_color}{\text{TPR} = \frac{\text{True Positives}}{(\text{True Positives})+(\text{False Negatives})}}
\end{equation}
\vspace{-0.2in}
\begin{equation}
    \textcolor{new_color}{\text{FPR} = \frac{\text{False Positives}}{(\text{False Positives})+(\text{True Negatives})}}
\end{equation}
ROC curves can allow us to choose a threshold where the TPR$\rightarrow1$ while the FPR$\rightarrow0$. For a perfect classifier (TPR$=1$, FPR$=0$), the ``elbow" of the ROC curve tightly fits in the upper left corner. This perfect classifier is plotted in each of the subplots of \textit{Figure \ref{fig:ROC}} along with the ROC curves of the machine learning methods. We can also compute the area under the curve (AUC) for each of these ROC curves. AUC is a threshold-sensitive measure of model accuracy. A perfect classifier has an AUC $=1$ whereas a completely random classifier would have an AUC $=0.5$. 

We listed the AUC values for each XRB class and each ML model in \textit{Figure \ref{fig:ROC}}. The AUC values for column 3 are the highest, ranging from 0.97 to 0.98, meaning that all three algorithms are nearly perfect classifiers for pulsars. For NPNS, the SVM model outperforms the others with an AUC $=0.9$ compared to AUC $=0.88$ for the KNN and BGP models. Across BHs, the differences between model accuracy for the methods are marginally smaller (BGP AUC $=0.86$, SVM AUC $=0.86$, KNN AUC $=0.85$).

Users of our software may use these ROC curves to choose an appropriate threshold for their own specific purposes. If a user's primary concern is reducing the FPR to zero, a higher probability threshold could be chosen based on the ROC curves. For instance, say a user primarily wants to minimize BH false positives with the KNN method, \textit{Figure \ref{fig:ROC}g} indicates that a probability threshold of $\sim$ 0.40 would keep the FPR at zero while providing them with an TPR $\sim$ 0.6. Depending on a user's exact purposes, they may instead consider a few false positives acceptable and would prefer to increase the TPR. In that type of scenario, a lower probability threshold may be more appropriate. Overall, these ROC curves allow users to choose a threshold that balances the FPR and TPR in a manner that is suitable for their own specific purposes.

\section{Computational Efficiency}
For a single full cross-validation run, the BGP requires $\sim$ 30 hours on the Lonestar 5 Supercomputer operated by Texas Advanced Computing Center (TACC); both the KNN and SVM can run on a personal computer (MacBook Pro 2.5 GHz Intel Core i7) with KNN taking about $\sim$ 10 minutes for one run of the entire cross-validation set and SVM taking about $\sim$  20 minutes.

\section{Discussion and Conclusion}
This paper compares three ML methods (BGP, KNN, SVM) to test their feasibility for accurate classification of the nature of the compact object in XRBs.  All three have relatively high predictive accuracy, ranging from 81.8$\%$ to 84.1$\%$ on average, with class-specific breakdowns as shown in \textit{ Figure \ref{fig:summary}}. In terms of predictive accuracy, the KNN model outperforms the SVM and BGP. The KNN model correctly predict the classifications of 84.1$\%$ of sources compared to 81.8$\%$ for the SVM and BGP. These differences in predictive power primarily stem from differences in misclassifications or indeterminate classifications of BHs across methods.

In terms of computational efficiency, the KNN and SVM model outperform the BGP. Specifically, the BGP requires $\sim 30-130$ hours on a supercomputer cluster whereas SVM and KNN can be run on a MacBook Pro (2.5 GHz Intel Core i7) within 10-120 minutes. 

In previous work (GP15), the largest number of sources mis-classified were LMNS bursters with only 33.3$\%$ correctly classified. In GP15, Bursters were excluded from the training set and only used for validation. We found that by including Bursters in the training set, we can achieve a significant improvement where now 100.0$\%$ of Bursters are correctly classified for the BGP, KNN and SVM model. However, the similarity between BHs and Bursters in quiescence leads to mis-classification of some BHs as Bursters.

To improve the prediction accuracy of the BGP classification method compared to GP15, we also increased the sampling of the data from 10$\%$ to 20$\%$ and perform 44-source cross-validation on the Texas Advanced Computing Center (TACC) Supercomputer. In future work, including larger samples of classified XRB sources for training could improve model performance and provide robust methods for classifying new sources, potentially even extragalactic sources from Chandra X-ray Observatory. In addition, these methods may also be applicable to Cataclysmic Variables in future studies. Adding dimensions to include other attributions that are easily available from the lightcurve dataset we used (such as long-term temporal behavior), could also significantly increase the predictive accuracy.

This work suggests that applying ML methods for understanding X-ray binary systems could potentially pave the way towards solutions to time-consuming problems such as measuring compact object masses, determining luminosity and strengths of magnetic fields, and the spatial distribution of XRBs in galaxies. This research provides the astrophysics community with efficient tools to classify novel X-ray binary systems, accelerating our understanding of XRBs' unique role in galaxy formation and evolution. We have made the software for all three methods publicly available on github\footnote{\url{https://github.com/zdebeurs/3ML_methods_for_XRB_classification/blob/main/Figures/README.md}}.






\subsection{Animations}

Animated three-dimensional versions of \textit{Figures \ref{fig:CCI}, \ref{fig:bhs_as_bursters}, \ref{fig:cygx3}, and \ref{fig:4u1700}} are available online$^3$. Each of these figures rotate from $0^{\circ}$ to $360^{\circ}$ and illustrate the spatial patterns in the data.


\vspace{0.1in}
\section*{Acknowledgements}

We thank Jonathan McDowell for his insightful knowledge of C that was instrumental in resolving package installation errors. We also thank Matt Ashby for his constructive and invaluable feedback in the writing process. Further, we thank Vinay Kashyap, Aneta Siemiginowksa and Josh Speagle for their statistical and machine learning expertise that furthered the work. We thank Alyssa Goodman for use of the Harvard Odyssey Supercomputer Cluster. We thank Adam Kraus and Andrew Vanderburg for use of the Texas Advanced Computing Center (TACC).

The SAO REU program is funded in part by the National Science Foundation REU and Department of Defense ASSURE programs under NSF Grant no.\ AST-1852268, and by the Smithsonian Institution.

This material is based upon work supported by the National Science Foundation Graduate Research Fellowship under Grant No. 1745302.

This research has made use of MAXI data provided by RIKEN, JAXA and the MAXI team.

The BGP computations presented in this paper were run on the \href{http://www.tacc.utexas.edu}{Texas Advanced Computing Center (TACC) at The University of Texas at Austin}. Thus, the authors acknowledge the TACC for providing HPC resources that have contributed to the research results reported within this paper. Some of the preliminary computations of the BGP were also run on the FASRC Odyssey cluster supported by the FAS Division of Science Research Computing Group at Harvard University.


\software{R \citep{R}, Kernlab \citep{kernlab}, ggplot2 \citep{ggplot2}, LIBSVM library \citep{CC01a}, class package \citep{classR}.
          }



\bibliography{bibliography}{}
\bibliographystyle{aasjournal}

\section*{\small{Appendix}}
\renewcommand{\thefigure}{A\arabic{figure}}
\setcounter{figure}{0}

\begin{figure*}[h!tp]
\centering
\includegraphics[clip,width=1.04\textwidth]{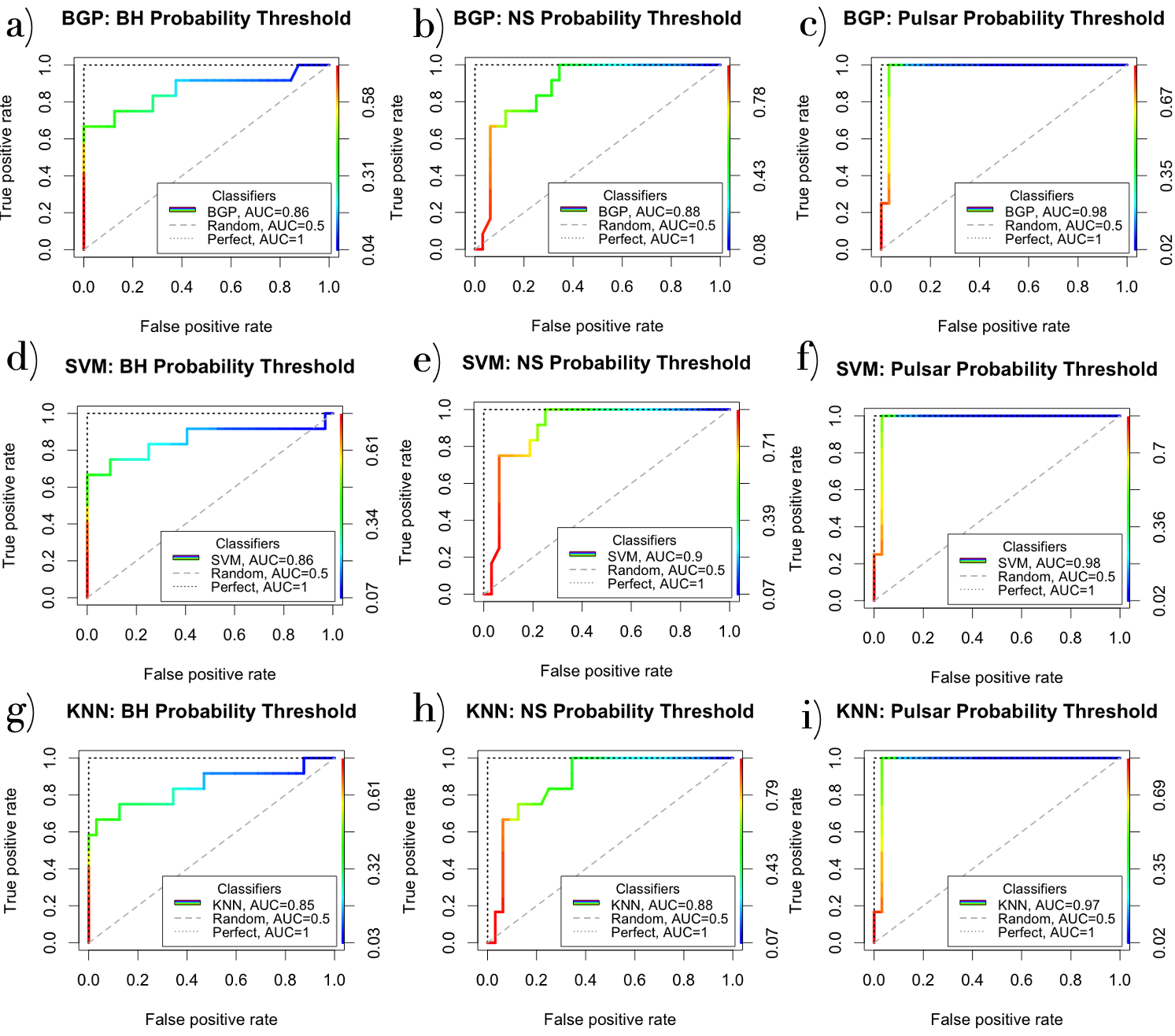}
\caption{\footnotesize{Receiver Operator Curves (ROC) for all three ML methods. Each column of subplots evaluates the performance for a single XRB class. Each rows of subplots focuses on performance of a single ML method across all three XRB classes.  Column I (a, d, g): ROC curves for various BH probability thresholds across ML methods. Column II (b, e, h): ROC curves for various NPNS probability thresholds across ML methods. Column III (c, f, i): ROC curves for various pulsar probability thresholds across ML methods. In each subplot, the dotted line represent a perfect classifier (AUC=1). The dashed line in each subplot represents a random classifier (AUC=0.5). The rainbow lines represent our actual classifiers where the color corresponds to a specific probability threshold as indicated in the colorbar on the right in each subplot.}
\label{fig:ROC}} 
\end{figure*}

\begin{deluxetable*}{ccCrlccccc}[ht!]
\tabletypesize{\footnotesize}
\tablecaption{BGP Probabilities and Predictions for Best Model (kernel=Laplace)
\label{table:A1}}
\tablecolumns{10}
\tablenum{A1}
\tablewidth{0pt}
\tablehead{
\colhead{System} &
\colhead{Prob(BH)*}&
\colhead{\ensuremath{\rm{SD^{\dagger}}}} &
\colhead{Prob(NPNS)} &
\colhead{SD}&
\colhead{Prob(Pulsar)}&
\colhead{SD}&
\colhead{Pred.}&
\colhead{Class}&
\colhead{Correct?}
}
\startdata
LMCX-3 & 0.67 & \num{2.4E-02} & 0.30 & \num{2.6E-02} & 0.03 & \num{3.3E-03} & BH & BH & Yes \\
LMCX-1 & 0.70 & \num{1.3E-02} & 0.24 & \num{1.2E-02} & 0.07 & \num{4.3E-03} & BH & BH & Yes \\
MAXIJ1535-571 & 0.58 & \num{1.3E-02} & 0.40 & \num{1.3E-02} & 0.03 & \num{2.0E-03} & BH & BH & Yes \\
4U1630-47& 0.43 & \num{3.1E-02} & 0.44 & \num{2.2E-02} & 0.13 & \num{1.0E-02} & BH or NPNS & BH & \textbf{No} \\
GX339-4 & 0.70 & \num{2.1E-02} & 0.24 & \num{1.7E-02} & 0.06 & \num{6.2E-03} & BH & BH & Yes \\
GRS1739-278& 0.27 & \num{8.3E-03} & 0.66 & \num{1.0E-02} & 0.07 & \num{7.0E-03} & NPNS & BH & \textbf{No} \\
H1743-322& 0.34 & \num{1.6E-02} & 0.49 & \num{1.4E-02} & 0.16 & \num{7.1E-03} & NPNS & BH & \textbf{No} \\
MAXIJ1820+070 & 0.45 & \num{1.0E-02} & 0.44 & \num{1.0E-02} & 0.11 & \num{6.0E-03} & BH or NPNS & BH & \textbf{No} \\
GRS1915 & 0.18 & \num{1.0E-02} & 0.68 & \num{2.0E-02} & 0.12 & \num{1.3E-02} & NPNS & BH & \textbf{No} \\
CygX-1 & 0.67 & \num{1.2E-02} & 0.29 & \num{1.0E-02} & 0.04 & \num{1.8E-03} & BH & BH & Yes \\
4U1957+115 & 0.45 & \num{1.1E-02} & 0.51 & \num{8.9E-03} & 0.03 & \num{3.3E-03} & NPNS & BH & \textbf{No} \\
CygX-3 & 0.07 & \num{1.1E-02} & 0.21 & \num{5.7E-03} & 0.72 & \num{1.1E-02} & Pulsar & BH & \textbf{No} \\
\hline
H0614+091 & 0.42 & \num{1.5E-02} & 0.54 & \num{1.8E-02} & 0.04 & \num{3.0E-03} & NPNS & NPNS & Yes \\
4U1254-690 & 0.36 & \num{1.6E-02} & 0.52 & \num{2.9E-02} & 0.12 & \num{1.4E-02} & NPNS & NPNS & Yes \\
CirX-1 & 0.32 & \num{1.9E-03} & 0.52 & \num{9.0E-03} & 0.17 & \num{9.0E-03} & NPNS & NPNS & Yes \\
4U1608-52 & 0.34 & \num{7.1E-03} & 0.62 & \num{6.6E-03} & 0.05 & \num{3.4E-03} & NPNS & NPNS & Yes \\
ScoX-1 & 0.04 & \num{2.4E-03} & 0.94 & \num{3.4E-03} & 0.02 & \num{1.8E-03} & NPNS & NPNS & Yes \\
H1636-536 & 0.29 & \num{7.9E-03} & 0.60 & \num{1.1E-02} & 0.10 & \num{4.5E-03} & NPNS & NPNS & Yes \\
4U1700-37 & 0.27 & \num{6.6E-03} & 0.34 & \num{2.4E-02} & 0.38 & \num{2.0E-02} & Pulsar & NPNS & \textbf{No} \\
GX349+2 & 0.04 & \num{3.4E-03} & 0.89 & \num{6.9E-03} & 0.07 & \num{6.0E-03} & NPNS & NPNS & Yes \\
4U1705-44 & 0.11 & \num{5.3E-03} & 0.84 & \num{8.7E-03} & 0.06 & \num{3.5E-03} & NPNS & NPNS & Yes \\
GX9+9 & 0.07 & \num{3.9E-03} & 0.89 & \num{4.4E-03} & 0.04 & \num{1.8E-03} & NPNS & NPNS & Yes \\
GX3+1 & 0.38 & \num{1.2E-02} & 0.50 & \num{1.1E-02} & 0.11 & \num{1.0E-02} & NPNS & NPNS & Yes \\
GX5-1 & 0.16 & \num{2.0E-02} & 0.79 & \num{2.0E-02} & 0.05 & \num{1.4E-03} & NPNS & NPNS & Yes \\
GX9+1 & 0.07 & \num{9.1E-03} & 0.88 & \num{9.6E-03} & 0.04 & \num{1.3E-03} & NPNS & NPNS & Yes \\
GX13+1 & 0.16 & \num{2.2E-02} & 0.81 & \num{2.3E-02} & 0.03 & \num{2.2E-03} & NPNS & NPNS & Yes \\
GX17+2 & 0.04 & \num{5.5E-03} & 0.91 & \num{9.4E-03} & 0.05 & \num{5.1E-03} & NPNS & NPNS & Yes \\
SerX-1 & 0.11 & \num{6.9E-03} & 0.85 & \num{6.7E-03} & 0.04 & \num{3.7E-03} & NPNS & NPNS & Yes \\
HETEJ1900.1-2455 & 0.28 & \num{7.0E-03} & 0.52 & \num{4.4E-03} & 0.20 & \num{1.0E-02} & NPNS & NPNS & Yes \\
AqlX-1 & 0.31 & \num{5.5E-03} & 0.65 & \num{5.6E-03} & 0.04 & \num{4.2E-03} & NPNS & NPNS & Yes \\
4U1916-053 & 0.41 & \num{1.0E-02} & 0.48 & \num{3.9E-03} & 0.11 & \num{9.1E-03} & NPNS & NPNS & Yes \\
CygX-2 & 0.15 & \num{9.3E-03} & 0.82 & \num{1.1E-02} & 0.03 & \num{2.6E-03} & NPNS & NPNS & Yes \\
\hline
SMCX-1 & 0.13 & \num{1.7E-02} & 0.23 & \num{1.0E-02} & 0.65 & \num{2.1E-02} & Pulsar & Pulsar & Yes \\
LMCX-4 & 0.19 & \num{1.3E-02} & 0.34 & \num{1.1E-02} & 0.46 & \num{1.5E-02} & Pulsar & Pulsar & Yes \\
1A0535+262 & 0.16 & \num{1.1E-02} & 0.32 & \num{1.1E-02} & 0.53 & \num{1.7E-02} & Pulsar & Pulsar & Yes \\
VelaX-1 & 0.25 & \num{1.5E-02} & 0.15 & \num{1.1E-02} & 0.60 & \num{1.6E-02} & Pulsar & Pulsar & Yes \\
GROJ1008-57 & 0.18 & \num{5.8E-03} & 0.28 & \num{8.6E-03} & 0.53 & \num{9.7E-03} & Pulsar & Pulsar & Yes \\
CenX-3 & 0.14 & \num{6.7E-03} & 0.17 & \num{3.1E-03} & 0.69 & \num{7.3E-03} & Pulsar & Pulsar & Yes \\
GX301-2 & 0.15 & \num{1.3E-02} & 0.08 & \num{4.3E-03} & 0.77 & \num{1.5E-02} & Pulsar & Pulsar & Yes \\
4U1538-52 & 0.13 & \num{3.5E-03} & 0.23 & \num{1.1E-02} & 0.64 & \num{1.4E-02} & Pulsar & Pulsar & Yes \\
4U1626-67 & 0.12 & \num{8.1E-03} & 0.25 & \num{1.8E-02} & 0.63 & \num{2.3E-02} & Pulsar & Pulsar & Yes \\
HerX-1 & 0.09 & \num{3.3E-03} & 0.10 & \num{5.7E-03} & 0.82 & \num{7.9E-03} & Pulsar & Pulsar & Yes \\
OAO1657 & 0.16 & \num{6.4E-03} & 0.24 & \num{1.1E-02} & 0.60 & \num{1.7E-02} & Pulsar & Pulsar & Yes \\
4U1822-37 & 0.10 & \num{5.9E-03} & 0.12 & \num{8.5E-03} & 0.77 & \num{1.0E-02} & Pulsar & Pulsar & Yes
\enddata
\tablenotetext{*}{\footnotesize{The Prob(X) is the median probability across 5 runs of belonging to class X, where X can be BH, NPNPNS, or pulsar.}}
\tablenotetext{\dagger}{\footnotesize{These errors are the standard deviations across 5 runs of the best BGP model. These errors should be interpreted as an indication of variation across runs.}}
\end{deluxetable*}


\begin{deluxetable*}{ccCrlccccc}[h!]
\tabletypesize{\footnotesize}
\tablecaption{KNN Probabilities and Predictions for Best Model (k=24)
\label{table:A2}}
\tablecolumns{10}
\tablenum{A2}
\tablewidth{0pt}
\tablehead{
\colhead{System} &
\colhead{Prob(BH)*}&
\colhead{\ensuremath{\rm{SD^{\dagger}}}} &
\colhead{Prob(NPNS)} &
\colhead{SD}&
\colhead{Prob(Pulsar)}&
\colhead{SD}&
\colhead{Pred.}&
\colhead{Class}&
\colhead{Correct?}
}
\startdata
LMCX-3 & 0.67 & \num{2.1E-02} & 0.31 & \num{2.2E-02} & 0.02 & \num{1.9E-03} & BH & BH & Yes \\
LMCX-1 & 0.71 & \num{1.9E-02} & 0.25 & \num{1.7E-02} & 0.05 & \num{4.5E-03} & BH & BH & Yes \\
MAXIJ1535-571 & 0.56 & \num{1.5E-02} & 0.42 & \num{1.4E-02} & 0.02 & \num{2.0E-03} & BH & BH & Yes \\
4U1630-47& 0.42 & \num{2.7E-02} & 0.42 & \num{2.7E-02} & 0.15 & \num{1.9E-02} & BH or NPNS & BH & \textbf{No} \\
GX339-4 & 0.73 & \num{2.3E-02} & 0.23 & \num{1.7E-02} & 0.04 & \num{9.0E-03} & BH & BH & Yes \\
GRS1739-278& 0.24 & \num{1.1E-02} & 0.69 & \num{1.4E-02} & 0.06 & \num{1.0E-02} & NPNS & BH & \textbf{No} \\
H1743-322& 0.32 & \num{1.4E-02} & 0.50 & \num{1.5E-02} & 0.17 & \num{1.1E-02} & NPNS & BH & \textbf{No} \\
MAXIJ1820+070 & 0.47 & \num{9.7E-03} & 0.44 & \num{9.1E-03} & 0.10 & \num{7.9E-03} & BH & BH & Yes \\
GRS1915 & 0.15 & \num{1.5E-02} & 0.71 & \num{2.0E-02} & 0.13 & \num{1.6E-02} & NPNS & BH & \textbf{No} \\
CygX-1 & 0.68 & \num{1.5E-02} & 0.29 & \num{1.5E-02} & 0.02 & \num{3.3E-03} & BH & BH & Yes \\
4U1957+115 & 0.44 & \num{1.6E-02} & 0.53 & \num{1.6E-02} & 0.03 & \num{4.0E-03} & NPNS & BH & \textbf{No} \\
CygX-3 & 0.06 & \num{1.0E-02} & 0.21 & \num{1.6E-02} & 0.74 & \num{1.6E-02} & Pulsar & BH & \textbf{No} \\
\hline
H0614+091 & 0.38 & \num{1.8E-02} & 0.58 & \num{2.0E-02} & 0.03 & \num{3.8E-03} & NPNS & NPNS & Yes \\
4U1254-690 & 0.34 & \num{2.2E-02} & 0.54 & \num{2.9E-02} & 0.12 & \num{1.4E-02} & NPNS & NPNS & Yes \\
CirX-1 & 0.29 & \num{1.2E-02} & 0.52 & \num{1.3E-02} & 0.18 & \num{1.6E-02} & NPNS & NPNS & Yes \\
4U1608-52 & 0.31 & \num{1.4E-02} & 0.64 & \num{1.5E-02} & 0.05 & \num{4.8E-03} & NPNS & NPNS & Yes \\
ScoX-1 & 0.03 & \num{8.3E-03} & 0.95 & \num{1.0E-02} & 0.02 & \num{4.7E-03} & NPNS & NPNS & Yes \\
H1636-536 & 0.28 & \num{1.5E-02} & 0.61 & \num{1.7E-02} & 0.10 & \num{5.1E-03} & NPNS & NPNS & Yes \\
4U1700-37 & 0.25 & \num{1.3E-02} & 0.34 & \num{1.9E-02} & 0.41 & \num{2.1E-02} & Pulsar & NPNS & \textbf{No} \\
GX349+2 & 0.03 & \num{6.3E-03} & 0.93 & \num{8.7E-03} & 0.04 & \num{4.7E-03} & NPNS & NPNS & Yes \\
4U1705-44 & 0.10 & \num{5.9E-03} & 0.86 & \num{9.3E-03} & 0.04 & \num{4.8E-03} & NPNS & NPNS & Yes \\
GX9+9 & 0.04 & \num{1.0E-02} & 0.94 & \num{1.4E-02} & 0.02 & \num{4.0E-03} & NPNS & NPNS & Yes \\
GX3+1 & 0.41 & \num{1.7E-02} & 0.48 & \num{1.8E-02} & 0.10 & \num{9.3E-03} & NPNS & NPNS & Yes \\
GX5-1 & 0.13 & \num{1.8E-02} & 0.83 & \num{1.9E-02} & 0.03 & \num{8.1E-03} & NPNS & NPNS & Yes \\
GX9+1 & 0.07 & \num{1.1E-02} & 0.90 & \num{1.4E-02} & 0.03 & \num{6.6E-03} & NPNS & NPNS & Yes \\
GX13+1 & 0.13 & \num{1.7E-02} & 0.85 & \num{1.9E-02} & 0.02 & \num{3.9E-03} & NPNS & NPNS & Yes \\
GX17+2 & 0.03 & \num{5.5E-03} & 0.94 & \num{8.7E-03} & 0.03 & \num{4.2E-03} & NPNS & NPNS & Yes \\
SerX-1 & 0.09 & \num{7.3E-03} & 0.89 & \num{7.2E-03} & 0.02 & \num{2.5E-03} & NPNS & NPNS & Yes \\
HETEJ1900.1-2455 & 0.27 & \num{1.1E-02} & 0.54 & \num{2.0E-02} & 0.19 & \num{1.3E-02} & NPNS & NPNS & Yes \\
AqlX-1 & 0.30 & \num{1.2E-02} & 0.67 & \num{1.2E-02} & 0.03 & \num{4.1E-03} & NPNS & NPNS & Yes \\
4U1916-053 & 0.43 & \num{1.5E-02} & 0.46 & \num{8.7E-03} & 0.11 & \num{1.2E-02} & NPNS & NPNS & Yes \\
CygX-2 & 0.12 & \num{1.3E-02} & 0.86 & \num{1.6E-02} & 0.03 & \num{4.3E-03} & NPNS & NPNS & Yes \\
\hline
SMCX-1 & 0.13 & \num{1.5E-02} & 0.25 & \num{1.5E-02} & 0.63 & \num{2.3E-02} & Pulsar & Pulsar & Yes \\
LMCX-4 & 0.20 & \num{1.2E-02} & 0.36 & \num{1.6E-02} & 0.44 & \num{1.4E-02} & Pulsar & Pulsar & Yes \\
1A0535+262 & 0.17 & \num{1.8E-02} & 0.30 & \num{1.1E-02} & 0.53 & \num{2.0E-02} & Pulsar & Pulsar & Yes \\
VelaX-1 & 0.30 & \num{1.9E-02} & 0.15 & \num{1.4E-02} & 0.55 & \num{2.3E-02} & Pulsar & Pulsar & Yes \\
GROJ1008-57 & 0.17 & \num{8.9E-03} & 0.29 & \num{7.5E-03} & 0.54 & \num{9.6E-03} & Pulsar & Pulsar & Yes \\
CenX-3 & 0.12 & \num{1.3E-02} & 0.18 & \num{1.0E-02} & 0.69 & \num{1.3E-02} & Pulsar & Pulsar & Yes \\
GX301-2 & 0.22 & \num{2.2E-02} & 0.07 & \num{9.3E-03} & 0.71 & \num{1.9E-02} & Pulsar & Pulsar & Yes \\
4U1538-52 & 0.12 & \num{1.0E-02} & 0.24 & \num{1.6E-02} & 0.63 & \num{1.7E-02} & Pulsar & Pulsar & Yes \\
4U1626-67 & 0.11 & \num{8.5E-03} & 0.28 & \num{1.9E-02} & 0.60 & \num{2.0E-02} & Pulsar & Pulsar & Yes \\
HerX-1 & 0.08 & \num{5.7E-03} & 0.09 & \num{8.2E-03} & 0.82 & \num{8.6E-03} & Pulsar & Pulsar & Yes \\
OAO1657 & 0.14 & \num{1.0E-02} & 0.25 & \num{1.2E-02} & 0.61 & \num{2.1E-02} & Pulsar & Pulsar & Yes \\
4U1822-37 & 0.09 & \num{8.1E-03} & 0.11 & \num{1.3E-02} & 0.79 & \num{1.3E-02} & Pulsar & Pulsar & Yes
\enddata
\tablenotetext{*}{The Prob(X) is the median probability across 10 runs of belonging to class X, where X can be BH, NPNS, or pulsar.}
\tablenotetext{\dagger}{These errors are the standard deviations across ten runs of the best value of $k$. These errors should be interpreted as an indication of variation across runs.}
\end{deluxetable*}

\begin{deluxetable*}{ccCrlccccc}[ht!]
\tabletypesize{\footnotesize}
\tablecaption{SVM Probabilities and Predictions for Best Model (C = 0.655, gamma = 0.585)
\label{table:A3}}
\tablecolumns{10}
\tablenum{A3}
\tablewidth{0pt}
\tablehead{
\colhead{System} &
\colhead{Prob(BH)*}&
\colhead{\ensuremath{\rm{SD^{\dagger}}}} &
\colhead{Prob(NPNS)} &
\colhead{SD}&
\colhead{Prob(Pulsar)}&
\colhead{SD}&
\colhead{Pred.}&
\colhead{Class}&
\colhead{Correct?}
}
\startdata
LMCX-3 & 0.70 & \num{1.9E-02} & 0.26 & \num{2.0E-02} & 0.03 & \num{2.7E-03} & BH & BH & Yes \\
LMCX-1 & 0.72 & \num{1.5E-02} & 0.21 & \num{1.6E-02} & 0.07 & \num{6.7E-03} & BH & BH & Yes \\
MAXIJ1535-571 & 0.59 & \num{2.9E-02} & 0.35 & \num{2.8E-02} & 0.06 & \num{1.6E-03} & BH & BH & Yes \\
4U1630-47& 0.43 & \num{6.5E-02} & 0.47 & \num{5.6E-02} & 0.10 & \num{2.0E-02} & BH or NPNS & BH & \textbf{No} \\
GX339-4 & 0.67 & \num{2.8E-02} & 0.27 & \num{2.3E-02} & 0.06 & \num{7.5E-03} & BH & BH & Yes \\
GRS1739-278& 0.24 & \num{1.5E-02} & 0.69 & \num{1.5E-02} & 0.06 & \num{8.4E-03} & NPNS & BH & \textbf{No} \\
H1743-322& 0.31 & \num{1.7E-02} & 0.53 & \num{1.7E-02} & 0.15 & \num{9.6E-03} & NPNS & BH & \textbf{No} \\
MAXIJ1820+070 & 0.41 & \num{1.1E-02} & 0.49 & \num{1.1E-02} & 0.10 & \num{7.4E-03} & NPNS & BH & \textbf{No} \\
GRS1915 & 0.20 & \num{1.5E-02} & 0.72 & \num{1.9E-02} & 0.08 & \num{8.2E-03} & NPNS & BH & \textbf{No} \\
CygX-1 & 0.67 & \num{1.4E-02} & 0.31 & \num{1.4E-02} & 0.03 & \num{1.5E-03} & BH & BH & Yes \\
4U1957+115 & 0.47 & \num{2.1E-02} & 0.49 & \num{2.1E-02} & 0.04 & \num{3.2E-03} & NPNS & BH & \textbf{No} \\
CygX-3 & 0.07 & \num{1.2E-02} & 0.19 & \num{1.5E-02} & 0.74 & \num{2.0E-02} & Pulsar & BH & \textbf{No} \\
\hline
H0614+091 & 0.37 & \num{2.1E-02} & 0.60 & \num{2.2E-02} & 0.04 & \num{3.7E-03} & NPNS & NPNS & Yes \\
4U1254-690 & 0.37 & \num{1.6E-02} & 0.53 & \num{2.2E-02} & 0.10 & \num{1.4E-02} & NPNS & NPNS & Yes \\
CirX-1 & 0.27 & \num{8.3E-03} & 0.56 & \num{1.1E-02} & 0.17 & \num{1.4E-02} & NPNS & NPNS & Yes \\
4U1608-52 & 0.26 & \num{1.5E-02} & 0.70 & \num{1.7E-02} & 0.04 & \num{2.5E-03} & NPNS & NPNS & Yes \\
ScoX-1 & 0.12 & \num{7.5E-03} & 0.84 & \num{8.6E-03} & 0.03 & \num{3.7E-03} & NPNS & NPNS & Yes \\
H1636-536 & 0.23 & \num{9.3E-03} & 0.68 & \num{1.1E-02} & 0.08 & \num{6.4E-03} & NPNS & NPNS & Yes \\
4U1700-37 & 0.22 & \num{1.1E-02} & 0.39 & \num{2.3E-02} & 0.40 & \num{1.8E-02} & NPNS or Pulsar & NPNS & \textbf{No} \\
GX349+2 & 0.11 & \num{8.8E-03} & 0.83 & \num{6.2E-03} & 0.06 & \num{5.2E-03} & NPNS & NPNS & Yes \\
4U1705-44 & 0.13 & \num{6.4E-03} & 0.82 & \num{9.8E-03} & 0.05 & \num{5.0E-03} & NPNS & NPNS & Yes \\
GX9+9 & 0.13 & \num{9.1E-03} & 0.83 & \num{1.1E-02} & 0.05 & \num{2.6E-03} & NPNS & NPNS & Yes \\
GX3+1 & 0.36 & \num{1.6E-02} & 0.56 & \num{1.6E-02} & 0.07 & \num{6.4E-03} & NPNS & NPNS & Yes \\
GX5-1 & 0.20 & \num{1.4E-02} & 0.77 & \num{1.4E-02} & 0.03 & \num{2.9E-03} & NPNS & NPNS & Yes \\
GX9+1 & 0.12 & \num{1.1E-02} & 0.84 & \num{1.3E-02} & 0.04 & \num{2.9E-03} & NPNS & NPNS & Yes \\
GX13+1 & 0.21 & \num{2.1E-02} & 0.77 & \num{2.2E-02} & 0.02 & \num{2.7E-03} & NPNS & NPNS & Yes \\
GX17+2 & 0.11 & \num{1.1E-02} & 0.84 & \num{1.5E-02} & 0.06 & \num{5.6E-03} & NPNS & NPNS & Yes \\
SerX-1 & 0.15 & \num{7.5E-03} & 0.81 & \num{8.6E-03} & 0.04 & \num{4.9E-03} & NPNS & NPNS & Yes \\
HETEJ1900.1-2455 & 0.26 & \num{1.4E-02} & 0.57 & \num{1.8E-02} & 0.18 & \num{1.3E-02} & NPNS & NPNS & Yes \\
AqlX-1 & 0.23 & \num{7.8E-03} & 0.73 & \num{9.5E-03} & 0.04 & \num{3.8E-03} & NPNS & NPNS & Yes \\
4U1916-053 & 0.26 & \num{1.1E-02} & 0.65 & \num{1.2E-02} & 0.09 & \num{6.5E-03} & NPNS & NPNS & Yes \\
CygX-2 & 0.20 & \num{1.0E-02} & 0.77 & \num{1.1E-02} & 0.03 & \num{1.8E-03} & NPNS & NPNS & Yes \\
\hline
SMCX-1 & 0.13 & \num{1.6E-02} & 0.18 & \num{1.4E-02} & 0.68 & \num{2.3E-02} & Pulsar & Pulsar & Yes \\
LMCX-4 & 0.19 & \num{1.7E-02} & 0.35 & \num{2.1E-02} & 0.47 & \num{2.3E-02} & Pulsar & Pulsar & Yes \\
1A0535+262 & 0.10 & \num{8.4E-03} & 0.33 & \num{2.0E-02} & 0.56 & \num{2.2E-02} & Pulsar & Pulsar & Yes \\
VelaX-1 & 0.26 & \num{2.9E-02} & 0.11 & \num{1.1E-02} & 0.63 & \num{3.0E-02} & Pulsar & Pulsar & Yes \\
GROJ1008-57 & 0.14 & \num{6.0E-03} & 0.27 & \num{1.0E-02} & 0.59 & \num{1.3E-02} & Pulsar & Pulsar & Yes \\
CenX-3 & 0.11 & \num{7.8E-03} & 0.14 & \num{8.2E-03} & 0.75 & \num{9.0E-03} & Pulsar & Pulsar & Yes \\
GX301-2 & 0.21 & \num{2.4E-02} & 0.07 & \num{6.4E-03} & 0.72 & \num{2.5E-02} & Pulsar & Pulsar & Yes \\
4U1538-52 & 0.11 & \num{5.8E-03} & 0.22 & \num{1.2E-02} & 0.67 & \num{1.4E-02} & Pulsar & Pulsar & Yes \\
4U1626-67 & 0.11 & \num{7.9E-03} & 0.25 & \num{1.7E-02} & 0.63 & \num{2.2E-02} & Pulsar & Pulsar & Yes \\
HerX-1 & 0.07 & \num{3.5E-03} & 0.08 & \num{7.8E-03} & 0.85 & \num{8.8E-03} & Pulsar & Pulsar & Yes \\
OAO1657 & 0.14 & \num{8.3E-03} & 0.23 & \num{1.6E-02} & 0.62 & \num{2.1E-02} & Pulsar & Pulsar & Yes \\
4U1822-37 & 0.10 & \num{7.0E-03} & 0.10 & \num{1.3E-02} & 0.80 & \num{1.5E-02} & Pulsar & Pulsar & Yes
\enddata
\tablenotetext{*}{The Prob(X) is the median probability across 10 runs of belonging to class X, where X can be BH, NPNS, or pulsar.}
\tablenotetext{\dagger}{These errors are the standard deviations across ten runs of the best SVM model. These errors should be interpreted as an indication of variation across runs.}
\end{deluxetable*}

\begin{figure*}[h!tp]
\epsscale{1.3}
\plotone{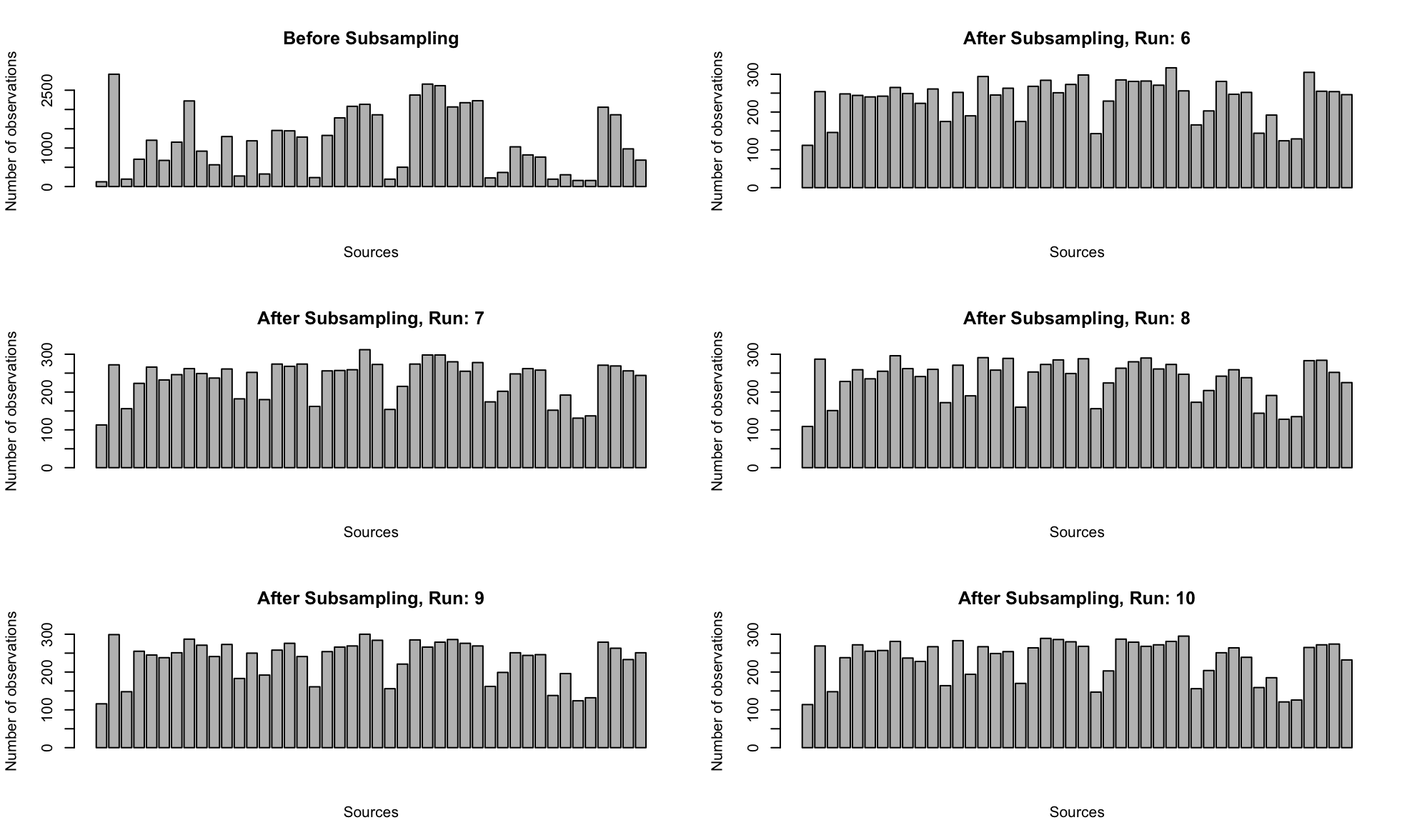}
\caption{\footnotesize{Distribution of the number of observations per source and five examples of  distributions after subsampling.  These and the additional five subsamples (Shown in \textit{Figure \ref{fig:subsamp}}) were used to test KNN and SVM runs.}
\label{fig:subsamp2}}
\end{figure*}

\end{document}